\documentclass{emulateapj}

\usepackage{bm}
\usepackage{amsmath}

\newcommand{\gcc}{\mbox{g cm$^{-3}$}}
\newcommand{\beq}{\begin{equation}}
\newcommand{\eeq}{\end{equation}}
\newcommand{\bea}{\begin{eqnarray}}
\newcommand{\eea}{\end{eqnarray}}
\newcommand{\req}[1]{Eq.\ (\ref{#1})}
\newcommand{\mel}{m_\mathrm{e}}
\newcommand{\mpr}{m_\mathrm{p}}
\newcommand{\mH}{m_\mathrm{H}}
\newcommand{\dd}{\mathrm{d}}
\newcommand{\am}{a_\mathrm{m}}
\newcommand{\omc}{\omega_\mathrm{ce}}
\newcommand{\omp}{\omega_\mathrm{cp}}
\newcommand{\kB}{k_\mathrm{B}}

\shorttitle{Atmospheres of magnetic neutron stars}

\begin{document}

\title{Equation of State and Opacities for 
       Hydrogen Atmospheres \\
      of Neutron Stars
        with Strong Magnetic Fields}

\author{Alexander Y. Potekhin}
\affil{Ioffe Physico-Technical Institute,
    Politekhnicheskaya 26, 194021 St.~Petersburg, Russia;
    \\
    Isaac Newton Institute of Chile, St.~Petersburg Branch, Russia}
\email{palex@astro.ioffe.rssi.ru}

\and

\author{Gilles Chabrier}
\affil{Ecole Normale Sup\'erieure de Lyon
    (CRAL, UMR CNRS No.\ 5574), % } \affil{
46 all\'ee d'Italie. 69364 Lyon Cedex 07, France}
\email{chabrier@ens-lyon.fr}

\slugcomment{Received 2002 June 19; accepted 2002 November 15}

% ***********************************************************
\begin{abstract}
We present an equation of state and radiative opacities 
for a strongly magnetized hydrogen plasma
at magnetic fields $B$, temperatures $T$, and densities 
$\rho$ typical 
for atmospheres of isolated neutron stars.
The first- and second-order thermodynamic functions,
monochromatic radiative opacities, and
Rosseland mean opacities are calculated 
and tabulated, taking account of partial ionization,
% in a wide density range
for $8\times10^{11}$ G $\leq B\leq 3\times10^{13}$ G, 
 $2\times10^5$ K $\leq T\leq 10^7$ K, 
 and a wide range of $\rho$.
We show that bound-bound and 
bound-free transitions give an important contribution to the 
opacities at $T \la (1$---$5)\times 10^6$ K in the considered range
of $B$ in the outer neutron-star atmosphere layers,
which may substantially modify the X-ray spectrum of a typical
magnetized neutron star.
In addition,
we re-evaluate opacities due to free-free transitions, 
taking into account the motion of both interacting particles,
electron and proton, in a strong magnetic field. 
Compared to the previous neutron-star atmosphere models,
the free-free absorption is strongly suppressed
 at photon frequencies below the proton cyclotron frequency.
 The latter result holds for any field strength,
 which prompts a revision of existing
 theoretical models of X-ray spectra of magnetar atmospheres.
\end{abstract}
\slugcomment{The Astrophysical Journal, 585, 955--974}
\setcounter{page}{955}
\maketitle

% ************************************************************
%                               TEXT BODY
% ************************************************************
\section{Introduction}
\label{sect-intro}
Models of neutron star atmospheres are needed for 
interpretation of their spectra and cooling. 
These atmospheres 
differ from the atmospheres of ordinary stars
because of the high gravity and magnetic fields
(for review, see, e.g., \citealp{Pavlov95,elounda}).

A magnetic field is called \emph{strong}\
if the electron cyclotron energy 
$\hbar\omc=\hbar eB/\mel c$
exceeds 1 a.u. -- i.e., the field strength $B$
is higher than $B_0=\mel^2 c\, e^3/\hbar^3
= 2.3505\times10^9$~G,
where $\mel$ is the electron mass, $e$ the elementary charge,
and $c$ the speed of light. Usually the field
 is called \emph{superstrong}
if $\hbar\omc > \mel c^2$,
that is $B > B_\mathrm{r}=\mel^2 c^3/e\hbar
= 4.414\times10^{13}$~G.
Most of the radio pulsars have
magnetic fields $B \sim 10^{12}$---$10^{13}$ G 
\citep*{tml93}, 
whereas anomalous
X-ray pulsars and soft gamma repeaters  
are thought to have superstrong fields (e.g., 
\citealp{mereghetti01,thompson00},
and references therein). 
Non-negligible amount of neutral atoms 
can exist in the photosphere
at typical neutron-star temperatures $T \sim 10^6$ K
(\citealp*{PCS99}, 
hereafter Paper I).
A strong magnetic field enhances atomic binding 
and makes the quantum-mechanical characteristics
of an atom dependent 
on its motion across the field 
(see \citealp{Lai01} for a recent review).
In photospheres of the neutron stars,
the field is, as a rule, \emph{strongly quantizing}, i.e., it sets
all the electrons on the ground Landau level. 
This occurs if $\beta_\mathrm{e} \gg 1$
and $\rho<\rho_B$,
where
\beq
   \beta_\mathrm{e} = \hbar\omc/\kB T
   \approx 134.3\,B_{12}/T_6 ,
\label{beta-e}
\eeq
 $\rho$ is the density, and
$\rho_B = \mH/(\pi^2\sqrt{2}\,\am^3)
\approx 7100\,B_{12}^{3/2}$ \gcc\ (for the hydrogen plasma).
Here and hereafter, $\mH = \mpr+\mel$,
$\mpr$ is the proton mass,  $\am=(\hbar c/eB)^{1/2}$
is the \emph{magnetic length},
 $\kB$ is the Boltzmann constant,
$B_{12}=B/10^{12}$~G, and $T_6=T/10^6$~K.

Opacities for the two polarization modes of radiation
are quite different in strongly magnetized plasmas (e.g., 
\citealp{Pavlov95},
and references therein),
which makes thermal emission of neutron stars polarized and
anisotropic \citep{Zavlin95}.
The mean opacities are strongly reduced at $\beta_\mathrm{e}\gg1$
(e.g., \citealp{silyak});
thus the bottom of the photosphere is shifted to high densities 
(e.g., \citealp{Pavlov95,LS97}).

The chemical composition of neutron-star atmospheres is not precisely
known.
Just after the neutron star birth in a supernova explosion,
the outer stellar envelope is most probably composed of iron. 
However, light elements
may be brought to the surface later (e.g., by fallback,
accretion, or encounters with comets).
Because of rapid gravitational sedimentation,
the lightest element will cover the surface
(see \citealp*{Brown02}). 
About $10^{12}$---$10^{14}$ grams
of hydrogen ($<10^{-19}$ M$_\odot$)
is sufficient to fill the entire photosphere. 

\citet{Shib92} presented the first 
model of hydrogen atmospheres with strong magnetic fields.
Later it was developed beyond the diffusion approximation 
\citep{SZ95} 
and used for astrophysical
predictions (e.g., \citealp{Zavlin95};
%GG% Pavlov et al.\ \citealp{Pavlov94};
\citealp*{Zane00,Zane01,HoLai,HoLai02,LaiHo02,Ozel01,Ozel02}) 
and for interpretation
of observed neutron-star spectra
         (e.g., \citealp*{PSZ95,PSZ96,Pavlov95,Ozel-ea}).

The above studies assume that the
atmosphere is fully ionized.
Meanwhile, it was recognized long ago 
(e.g., \citealp{Miller92})
that a significant contribution to the opacities 
of neutron-star photospheres with strong magnetic fields might
come from bound-bound and bound-free absorption by atoms.
Examples of monochromatic opacities 
in partially ionized iron
\citep*{RRM}
and hydrogen \citep*{PCS00}
 atmospheres
 confirmed this conjecture. 
In Paper I we have presented an equation of state (EOS) of a partially
ionized hydrogen plasma for the values of $T$ and $B$
typical for atmospheres of the radio pulsars.
Here we report results of extensive calculations
of thermodynamic functions based on the theory developed in Paper~I,
supplemented by calculations of the opacities
(monochromatic and Rosseland mean). Partial ionization and plasma 
nonideality are taken into account for 
$11.9 \leq \log_{10} B/\mathrm{G} \leq 13.5$ and 
$5.3 \leq \log_{10} T/\mathrm{K} \leq 7.0$.
Bound-bound and bound-free radiative transitions are treated
within the framework of a previously developed theory 
\citep{PP95,PP97}.
The free-free absorption cross sections are re-evaluated.
Whereas the previous authors considered 
photoabsorption by an electron scattered off a fixed Coulomb center,
we take into account the finite proton mass,
which has a nontrivial effect on the photoabsorption 
in a quantizing magnetic field.

The paper is composed as follows. 
In Sect.\ \ref{sect-input} we
formulate the main assumptions and give the basic formulae
used in our work.
Section \ref{sect-EOS} presents the EOS of partially ionized
hydrogen under conditions in neutron-star photospheres.
In Sect.\ \ref{sect-cross} we discuss various contributions
to the hydrogen photoabsorption cross sections 
in strong magnetic fields and derive a new formula for
the free-free cross section.
Opacities of hydrogen photospheres of the neutron stars
are discussed in Sect.\ \ref{sect-opac}.
Appendices give some detail of calculation of the free-free
cross sections.

% **************************************************************
%AA% \section{Basic equations and physics input}
      \section{Basic Equations and Physics Input}
\label{sect-input}
%
%AA% \subsection{Hydrogen atom moving in a magnetic field}
       \subsection{Hydrogen Atom Moving in a Magnetic Field}
\label{sect-H-atom}
If an atom rests without motion in a strong magnetic field,
there are two distinct classes of
its quantum states: at every value of the Landau quantum number $n$
and the magnetic quantum number $-s$ ($n\geq 0$, $s\geq -n$), 
there is one tightly bound
state, with
binding energy growing asymptotically as $[\ln (B/B_0)]^2$, 
and an infinite
series of hydrogenlike states 
with binding energies approaching
the energies of a field-free H atom (e.g.,
\citealp{CanutoVentura}). The atom is elongated:
its size along the magnetic field
$\bm{B}$ either decreases logarithmically (for 
the tightly bound states)
or remains nearly constant (for the hydrogenlike states), 
while the transverse radius is close to $\am$, decreasing as $B^{-1/2}$.
The radiative transition rates 
are different for the three basic polarizations:
the linear polarization along the field and the two circular
polarizations in the transverse plane. 

This simplicity is destroyed when atomic motion is taken into
account. 
The electric field, induced in the comoving 
frame of reference, breaks down the cylindrical symmetry.
In the nonrelativistic quantum mechanics, 
the binding energies and wave functions of the H atom are given 
by a solution of the Schr\"odinger equation with the 
two-particle Hamiltonian
\beq
  H = \frac{\pi_\mathrm{p}^2}{2{\mpr}} 
   + \frac{\pi_\mathrm{e}^2}{2\mel}
   - \frac{e^2}{|\bm{r}_\mathrm{e}-\bm{r}_\mathrm{p}|},
\eeq
where
$m_i$, $\bm{r}_i$, and $\bm{\pi}_i$
are the mass, radius, and kinetic momentum 
of the electron ($i=$e) or proton ($i=$p).
The kinetic momentum (related to the velocity) equals 
(e.g., \citealp{LaLi-QM})
\beq
   \bm{\pi}_i =
   \bm{p}_i -\frac{q_i}{c}\bm{A}(\bm{r}_i),
\label{pi_i}
\eeq
where $q_i$ is the charge of the $i$th particle
($q_\mathrm{e}=-q_\mathrm{p}=-e$),
$\bm{p}_i$ is the canonical momentum
(i.e., $\bm{p}_i=-\mathrm{i}\hbar\nabla_i$
in the coordinate representation), 
and $\bm{A}(\bm{r})$ is the vector potential of the field.
A conserved quantity related to the center-of-mass motion
is the \emph{pseudomomentum}
\beq
   \bm{K} = \bm{\pi}_\mathrm{p} + \bm{\pi}_\mathrm{e}
   +\frac{e}{c}\bm{B}\times(\bm{r}_\mathrm{p}-\bm{r}_\mathrm{e}).
\eeq
Let the $z$ axis of the Cartesian coordinates
$(x,y,z)$ be directed along $\bm{B}$.
Separating the center-of-mass motion
\citep*{GD,VDB92,P94}
and choosing the gauge of the vector potential in the form
\beq
   \bm{A}(\bm{r}) = \frac12 \bm{B} \times \left(\bm{r}
      - \frac{\mpr-\mel}{\mH}\, \bm{r}_0 \right),
\eeq
where $\bm{r}_0$ is arbitrary,
one comes to the effective one-particle Schr\"odinger equation
\beq
   \left( \frac{p_z^2}{2\mu}
   + H_\perp + H_{\bm{K}}(\bm{r}_0)
    - \frac{e^2}{|\bm{r}_0+\bm{r}|} \right)
   \psi(\bm{r}) = E \psi(\bm{r}).
\label{Schr}
\eeq
Here, $\bm{r}=\bm{r}_\mathrm{e}-\bm{r}_\mathrm{p}-\bm{r}_0$
is a ``shifted'' relative coordinate,
$p_z=-\mathrm{i}\hbar\,\partial/\partial z$ 
is the $z$ component of its conjugate momentum $\bm{p}$,
\bea&&
   H_\perp = \frac{\pi_\perp^2}{2\mu} 
   - \frac{e}{\mpr c} \bm{B}\cdot(\bm{r}\times\bm{p}),
\label{Hperp}
\\&&
   H_{\bm{K}}(\bm{r}_0) = \frac{1}{2\mH}
      \left(\bm{K}+\frac{e}{c}\bm{B}\times\bm{r}_0\right)^2
 \nonumber\\&&\hspace*{2em}
      + \frac{e}{\mH c}
      \left(\bm{K}+\frac{e}{c}\bm{B}\times\bm{r}_0\right)
      \cdot(\bm{B}\times\bm{r}),
\hspace*{2em}
\eea
and $\mu=\mel\mpr/\mH$ is the reduced mass.
In \req{Hperp},
\beq
   \bm{\pi} = \bm{p}+\frac{e}{2c}\bm{B}\times\bm{r},
\label{pi}
\eeq
and the subscript `$\perp$' denotes a vector component
perpendicular to $\bm{B}$.

 $H_{\bm{K}}(\bm{r}_0)$ turns to zero,
if we set $\bm{r}_0=\bm{r}_c$, where
\beq
   \bm{r}_c = \frac{c}{eB^2}\bm{B}\times\bm{K}
\eeq
is the relative guiding center (the difference
between the electron and proton guiding centers).
This choice of $\bm{r}_0$ is most useful for bound states
with large $K_\perp$ and the states of the continuum, whereas 
for bound states with small $K_\perp$ the choice $r_0=0$
is most appropriate \citep{P94,PP97}.

The eigenfunctions of $H_\perp$ 
are the Landau functions
$\Phi_{{n}s}(\bm{r}_\perp)$ (given, e.g., 
by Eq.\ (5) of \citet{PP93},
with eigenenergies 
\beq
   E_{{n}s}^\perp = {n}\hbar\omc 
   + ({n}+s) \hbar\omp ,
\eeq
where $\omp =(\mel/\mpr)\,\omc$
is the proton cyclotron frequency. 

It is convenient
to expand the wave function in the basis of $\Phi_{{n}s}(\bm{r}_\perp)$,
\beq
   \psi(\bm{r}) =  \sum_{{n}'s'} g_{{n}'s'}(z)
                        \Phi_{{n}'s'}(\bm{r}_\perp),
\label{expansion}
\eeq
and to label $\psi(\bm{r})$ by numbers ${n}$ and $s$, corresponding to
the leading term of this expansion.
The third quantum number $\nu$ then enumerates
``longitudinal'' energy levels.
The \emph{adiabatic approximation} widely used in the past
(e.g., \citealp{GD,CanutoVentura})
corresponds to retaining only one term $n'=n$, $s'=s$ in \req{expansion}.
We perform calculations without this approximation.

The total energy of the atom in \req{Schr} can be written as
\beq
   E= E_{{n}s}^\perp + E^\|_{{n}s\nu}(K_\perp).
\eeq
Here, the \emph{longitudinal energy} $E^\|_{{n}s\nu}(K_\perp)$ 
is negative for the bound and autoionizing (resonance) states
and positive for the continuum states, in which the motion
along $z$ is infinite (in the latter case, $\nu$ is continuous).
Since $n=0$ for the bound states of H atom in a strong magnetic field,
we will drop the number ${n}$ but imply $n=0$ for these states.
Then the binding energy is
\beq
   \epsilon_{s\nu}(K_\perp)=|E_{s\nu}^\|(K_\perp)|
       - s \hbar\omp .
\label{chi}
\eeq

The substitution of \req{expansion} in \req{Schr} reduces the problem
to the set of the \emph{coupled channel equations}
\bea&&
   (p_z^2/2\mu+E_{{n}'s'}^\perp - E)\, g_{{n}'s'}(z)
  \nonumber\\&&\hspace*{2em}
   + \sum_{{n}'',s''} V_{{n}''s'',{n}'s'}^{\mathrm{tot}}(z) g_{{n}''s''}(z)
   =0 ,
\hspace*{2em}\label{system}
\eea
where 
\beq
V_{{n}s,{n}'s'}^{\mathrm{tot}}(z)=
\langle {n}s|H_{\bm{K}}(\bm{r}_0)|{n}'s'\rangle_\perp + V_{{n}s,{n}'s'}(r_0,z)
\eeq 
is a total coupling potential, and
\beq
   V_{{n}s,{n}'s'}(r_0,z) = 
   \big\langle {n}s \big| -e^2/|\bm{r}_0+\bm{r}|\,\big|{n}'s' \big\rangle_\perp
\label{V}
\eeq
is an effective Coulomb potential.
Here, 
\beq
\langle {n}s|f(\bm{r})|{n}'s'\rangle_\perp
   = \int \Phi_{{n}s}^\ast(\bm{r}_\perp)  f(\bm{r})
    \Phi_{{n}'s'}(\bm{r}_\perp) \dd^2\bm{r}_\perp.
 \eeq

Numerical solutions of \req{Schr} for various $K_\perp$ were
presented by \citet{VDB92}. At superstrong fields, 
binding energies were calculated 
by \citet{LS95}.
The system of equations (\ref{system}) 
was numerically solved for the discrete
atomic states by \citet{P94}, and for the continuum
by \citet{PP97}. According to these studies,
an atom moving across the strong magnetic
field acquires a constant dipole moment
parallel to $\bm{r}_c$.
Those radiative transitions, which were dipole-forbidden 
for an atom at rest because of conservation 
of the $z$-projection of the angular momentum,
become allowed and should be taken into account 
in the atmosphere models.
If $K_\perp$ is small enough, the dipole moment is also small.
When $K_\perp$ exceeds a certain critical value, 
the atom becomes
\emph{decentered}: the average distance between the electron and proton
approaches $r_c$.
In this case, $K_\perp$
characterizes the electron-proton distance,
rather than the atomic velocity.
The binding energies (\ref{chi}) decrease with increasing $K_\perp$.
Asymptotically, at large $K_\perp$,
all longitudinal energies tend to $-e^2/r_c$. In this limit,
the cylindrical symmetry of the wave function and dipole
selection rules are restored, but the axis of symmetry is shifted
to the distance $r_c$ from the Coulomb center.

% ***********************************************************
%AA% \subsection{Thermodynamic model}
       \subsection{Thermodynamic Model}
\label{sect-thermomodel}
The EOS for partially ionized hydrogen in 
strong magnetic fields was constructed and discussed in
Paper I. We employ the free energy minimization technique
in the ``chemical picture'' of a plasma
(for discussion of its advantages and limitations
see, e.g., \citealp*{SCVH,P96b}).
The treatment is based on the framework of the free energy model
developed by \citet{SC91,SC92} at $B=0$ 
(see also \citealp{SCVH} and Sect.\ II of Paper I)
and extends it to the strong magnetic field case.

We consider a plasma composed of $N_\mathrm{p}$ protons, 
$N_\mathrm{e}$ electrons, $N_\mathrm{H}$
hydrogen atoms, and $N_\mathrm{mol}$ molecules in a volume $V$,
the number densities being $n_j\equiv N_j/V$.
The Helmholtz free energy is written as the sum
\beq
   F = F_\mathrm{id}^\mathrm{e} + F_\mathrm{id}^\mathrm{p} 
      + F_\mathrm{id}^\mathrm{neu}
       + F_\mathrm{ex}^\mathrm{C} + F_\mathrm{ex}^\mathrm{neu},
\label{Fren}
\eeq
where $F_\mathrm{id}^\mathrm{e}$, $F_\mathrm{id}^\mathrm{p}$, and
$F_\mathrm{id}^\mathrm{neu}$ are the free energies of ideal gases
of the electrons, protons, and neutral species, respectively, 
$F_\mathrm{ex}^\mathrm{C}$ takes into account the Coulomb plasma
nonideality, and $F_\mathrm{ex}^\mathrm{neu}$ is the nonideal
contribution which arises from interactions of bound species with
each other and with the electrons and protons. 
In \req{Fren} we have disregarded the additive contribution due
to photons, since it does not affect ionization equilibrium.
Moreover, generally we need not to assume thermodynamic
equilibrium of radiation with matter.
Ionization equilibrium is given by minimization of $F$ with
respect to particle numbers under the stoichiometric constraints,
provided the total number $N_0$ of protons (free and bound)
is fixed. The latter number is determined by the total mass
density:
$
  n_0 \equiv N_0/V \approx \rho/\mH.
$

The first term in \req{Fren} is
 $
   F_\mathrm{id}^\mathrm{e}  = 
     \mu_\mathrm{e} N_\mathrm{e} - P_\mathrm{e} V,
$
where $\mu_\mathrm{e}$ and $P_\mathrm{e}$ are
the chemical potential and pressure of the ideal Fermi gas,
respectively. 
They are obtained as functions of 
the electron number density
$n_\mathrm{e}$ and temperature $T$
from equations
(e.g., \citealp{bh82})
\bea&&\hspace*{-2em}
       P_\mathrm{e} =
   P_\mathrm{r}\,{b \tau_0^{3/2}\over\sqrt2\pi^2}
   \sum_{n=0}^\infty (2-\delta_{n0})
           (1+2bn)^{1/4} I_{1/2}(\chi_n,\tau_n),
\label{presmag}
\\&&\hspace*{-2em}
   n_\mathrm{e} =  \lambda_\mathrm{C}^{-3} \,{\tau_0 b \over 2\pi^2} 
  \sum_{n=0}^\infty (2-\delta_{n0})  
 \left[ \sqrt{2\over\tau_n} 
     {\partial\over \partial\chi_n} I_{1/2}(\chi_n,\tau_n)
  \right],
\label{densmag}
\eea
where 
\bea
&&\!\!
   I_{1/2}(\chi,\tau) = 
   \int_0^\infty (\mathrm{e}^{t-\chi}+1)^{-1}\,
   \sqrt{t(1+\tau t/2)}\,\mathrm{d}t, 
\nonumber
\\&&\!\!
   \chi_n = \frac{\mu_\mathrm{e}}{\kB T}
   +\tau_0^{-1}-\tau_n^{-1},
\quad
   \tau_n = {T/T_\mathrm{r}\over\sqrt{1+2bn}},
  \nonumber
\eea
and $b= (B/B_\mathrm{r})$. 
In these equations,
$\lambda_\mathrm{C}=\hbar/(\mel c)^2\approx
3.8616\times10^{-11}$ cm,
$P_\mathrm{r}=\mel c^2/\lambda_\mathrm{C}^3
\approx 1.4218\times10^{25}$ dyn cm$^{-2}$, and
$T_\mathrm{r}=\mel c^2/k_\mathrm{B}\approx5.930
\times10^9$~K
are the relativistic units of length, pressure, and temperature,
respectively, and $\mel c^2$ is \emph{not} included in 
$\mu_\mathrm{e}$.
We employ analytic fitting formulae to
the standard Fermi--Dirac integral
$I_{1/2}$ in \req{presmag}
and to the expression in square brackets in \req{densmag},
accurate within a few parts in $10^3$. These fits are
presented, respectively, 
in Sect.\ III of \citet{CP98}
and in Appendix C of \citet{P96a}.
When the electrons are nonrelativistic (as usually
 in the photospheres), Eqs.\ (\ref{presmag}) and
(\ref{densmag}) reproduce Eqs.\ (30) and (31) of Paper I.

The Coulomb free energy contribution consists of three parts, 
$F_\mathrm{ex}^\mathrm{C}=F_\mathrm{pp}+F_\mathrm{ee}+F_\mathrm{pe}$,
which represent, respectively, the proton-proton, electron-electron,
and proton-electron interactions. There was no detailed 
study of the influence of a strong magnetic field on these 
contributions in the $\rho$ and $T$ domain we are interested in.
Therefore we employ nonmagnetic expressions
\citep{CP98,PC00}, 
scaled with $B$. Specifically, the nonmagnetic expression
for $F_\mathrm{pe}$ is based on 
numerical results obtained in the 
hypernetted chain approximation for the linear
response theory with a local field correction 
\citep{C90}. The nonmagnetic expressions 
for $F_\mathrm{pp}$ and $F_\mathrm{ee}$
result from fitting the most accurate numerical 
results available in the literature 
 (see \citealp{PC00} for references).
In the strong magnetic field, the $B$-scaling of
the nonmagnetic $F_\mathrm{ex}^\mathrm{C}$ is 
constructed so as to match known low- and high-density limits
(Sect.\ IIIB of Paper I).

The free energy of nondegenerate and nonrelativistic
gas of protons, $F_\mathrm{id}^\mathrm{p}$,
is given by
\bea
   F_\mathrm{id}^\mathrm{p} / N_\mathrm{p} \kB T &=&
      \ln(2\pi \am^2\lambda_\mathrm{p} n_\mathrm{p})
    + \ln\left(1-\mathrm{e}^{-\beta_\mathrm{p}}\right)-1
 \nonumber\\&&
   +  \beta_\mathrm{p}/2
    - \ln\,[2\cosh( g_\mathrm{p}\beta_\mathrm{p}/4)] ,
 \eea
 where $g_\mathrm{p}=5.585$
is the proton gyromagnetic factor,
 $\lambda_\mathrm{p}=\hbar\,\sqrt{2\pi/(k_\mathrm{B}T \mpr)}$ 
is the proton thermal wavelength, and
$\beta_\mathrm{p} = \hbar\omp / \kB T
  \approx 0.0732\,B_{12}/T_6$.

Let $N_{s\nu}$ be the total number of H atoms with given 
quantum numbers $s$ and $\nu$ in the volume $V$,
and let $p_{s\nu}(K_\perp)\,\mathrm{d}^2 K_\perp$ be the
probability for such atom to have
a transverse pseudomomentum in an element $\mathrm{d}^2 K_\perp$
around $\bm{K}_\perp$. 
Then the ideal part of the free energy for hydrogen atoms
is 
\bea
&&\hspace*{-2em}
   F_\mathrm{id}^\mathrm{H} = k_\mathrm{B} T \sum_{s\nu} N_{s\nu}
    \int \Bigg\{
\ln\left[N_{s\nu} \lambda_\mathrm{H} {(2\pi\hbar)^2 \over V}
p_{s\nu}(K_\perp) \right]
 \nonumber\\&&
        - 1 - \epsilon_{s\nu}(K_\perp)/(k_\mathrm{B} T) \Bigg\}
       \, p_{s\nu}(K_\perp)\,\mathrm{d}^2 K_\perp
\nonumber\\[1ex]&&
   + N_\mathrm{H} \kB T \,\big\{ \beta_\mathrm{p}/2
    - \ln\,[2\cosh( g_\mathrm{p}\beta_\mathrm{p}/4)] \big\} ,
\label{FH1}
\eea
where $\lambda_\mathrm{H}\approx\lambda_\mathrm{p}$ 
is the thermal wavelength of an atom.
The probability density $p_{s\nu}(K_\perp)$ is calculated
in a thermodynamically consistent way 
from derivatives of the \emph{total} free energy $F$
with respect to the particle numbers. 
Molecules H$_2$ are treated in an approximate manner,
without taking into account their excited states and possible 
effects caused by their motion across the magnetic field
and rotation.
Finally, the nonideal part of the free energy of neutral
species, $F_\mathrm{ex}^\mathrm{neu}$, is obtained in frames of
the hard-sphere model, with effective radii depending on
the quantum numbers \emph{and pseudomomenta} of interacting
atoms (see Paper I for detail).

Once the free energy is obtained, its derivatives 
over $\rho$ and $T$ and their combinations provide
the other thermodynamic functions.

% ***********************************************************
\vspace*{5ex}
%AA% \subsection{Polarization modes and opacities: basic relations}
       \subsection{Polarization Modes and Opacities: Basic Relations}
\label{sect-opac-basic}
Propagation of radiation in magnetized plasmas
was discussed in many 
papers and monographs (e.g., \citealp{Ginzburg}).
At photon energies $\hbar\omega$ much higher than
\beq
   \hbar\omega_\mathrm{pl} =\left({4\pi\hbar^2 e^2 n_\mathrm{e} 
                      \over \mel} \right)^{1/2}
                      \approx 28.7\,\rho_0^{1/2}  \mathrm{~eV},
\label{plafreqe}
\eeq
where  $\omega_\mathrm{pl}$ is the electron plasma frequency
and $\rho_0\equiv\rho/\gcc$,
radiation propagates in the form of \emph{extraordinary}
(hereafter labeled by index $j=1$)
and \emph{ordinary} ($j=2$) normal modes.
These modes have different polarization vectors
$\bm{e}_j$
and different absorption and scattering
coefficients, which depend on the angle $\theta_B$
between the propagation direction and $\bm{B}$
(e.g., \citealp*{Kam82}).
The two modes interact with each other via scattering.
Vectors $\bm{e}_j$ for a fully ionized plasma
have been derived by \citet{Shafranov}.
\citet{Ventura79} gave
an instructive analysis of the plasma polarization modes
 relevant to the neutron stars.
\citet{GP73} formulated the radiative transfer
problem in terms of these modes.
They introduced the convenient real parameters $q$ and $p$,
which completely determine the normal mode polarization properties,
and which are defined as
\beq
   q+\mathrm{i}p=\frac{
   \varepsilon_{yy}-\varepsilon_{xx}\cos^2\theta_B
   +\varepsilon_{xz}\sin 2\theta_B
      -\varepsilon_{zz}\sin^2\theta_B
   }{
   2\,\mathrm{i}\,(\varepsilon_{xy}\cos\theta_B
   +\varepsilon_{yz}\sin\theta_B)},
\eeq
where $\varepsilon_{ij}$ are the components of the complex
permittivity tensor \citep{Ginzburg},
and the $z$ axis is directed along $\bm{B}$. The parameter $q$
determines the ellipticity of the normal modes,
and the parameter $p$ is associated with absorption of radiation.
In the most common case, one has 
\beq
   |q|\gg|p|,
 \qquad
  q(\omega,\theta_B)\approx \tilde{q}(\omega)\,
  \frac{\sin^2\theta_B}{2\cos\theta_B}.
\label{qp}
\eeq
These relations may be invalid in narrow frequency ranges
where resonant absorption occurs (e.g., near the 
electron or ion cyclotron resonance).

The formulae for $\bm{e}_j$ which take into account
contribution of the plasma ions, implied
in \citet{Shafranov},
have been explicitly written by
\citet{HoLai}. 
The electron-positron vacuum polarization in a strong magnetic field
dramatically changes the normal-mode properties
in certain $\rho$--$\omega$ domains
(see \citealp{PavlovGnedin}, for a review).
The vacuum dielectric tensor 
has been obtained by \citet{Adler}
at $B\ll B_\mathrm{r}$ and by \citet{HH97a,HH97b}
in both limits of $B\ll B_\mathrm{r}$ and $B\gg B_\mathrm{r}$.
Using these results,
\citet{HoLai02} derived convenient expressions
for the polarization vectors of normal modes, which take into account
the contributions from the electrons, ions, and vacuum.

The presence of bound species modifies the complex permittivity tensor 
and hence the properties of the normal modes. Their accurate treatment
in a partially ionized medium with a strong magnetic field
 is a complicated problem, which has not been solved yet. The
 normal polarization modes of a neutral gas of hydrogen atoms
 in strong magnetic fields
 were studied
 by \citet{BulikPavlov},
 who applied the Kramers--Kronig relations 
 to the bound-free and bound-bound
 atomic absorption coefficients
 obtained previously by \citet{PP93} 
 and \citet{PP95}.
 The thermal motion effects, which had not been calculated by that time
 for the bound-free transitions in strong magnetic fields, 
 were evaluated using a perturbation approximation \citep{PM93}.
The qualitative behavior of the polarization vectors
 proved to be the same as for the fully ionized plasma
 in a wide range of $\omega$ and $\theta_B$,
 where \req{qp} holds.
 However, there are quantitative differences.
 The frequencies at which \req{qp} is not valid,
 are of the order of the photoionization threshold $\omega_\mathrm{th}$
 and the principal bound-bound transition frequencies,
 where the resonant absorption takes place in the neutral gas
 (instead of $\omc$ and $\omp$ in the fully
 ionized plasma).
 The numerical values of $\tilde{q}(\omega)$ 
 are modified. 
 For instance, at $\omega$ well above  $\omega_\mathrm{th}$, 
 $\tilde{q}(\omega)$ is larger
 than it would be in the case of full ionization,
 which makes the normal mode polarization more linear.
 
Polarization properties of normal modes in a 
\emph{partially} ionized plasma with a strongly quantizing
magnetic field remain unexplored.
 Hereafter we consider mostly situations
 where the neutral fraction is small.
 Since even for completely nonionized gas 
 the properties of the polarization vectors (in particular,
 their dependence on $\theta_B$) are qualitatively the same
 as for the fully ionized plasma 
 (except for the resonant absorption frequency ranges),
 we assume that the polarization modes for the fully
 ionized plasma are a good approximation
 and adopt the formulae given by \citet{HoLai02}.

At a fixed photon frequency $\omega$,
the absorption opacity $\kappa_j^\mathrm{a}(\theta_B)$ 
in each mode $j$ and scattering opacities
$\kappa_{jj'}^\mathrm{s}(\theta_B)$ 
from mode $j$ into mode $j'$ can be
presented as (e.g., \citealp{Kam82})
\bea
&&\hspace*{-.7em}
   \kappa_j^\mathrm{a}(\theta_B) = \mH^{-1} \sum_{\alpha=-1}^1
     |e_{j,\alpha}(\theta_B)|^2 \,\sigma_\alpha^\mathrm{a},
\\&&\hspace*{-.7em}
 \kappa_{jj'}^\mathrm{s}(\theta_B) \!=\!\!
%     \left[ 
     {\frac34}
\!\!
  \sum_{\alpha=-1}^1 \!\!
     |e_{j,\alpha}(\theta_B)|^2 \,
  \nonumber\\&&\times
     {\sigma_\alpha^\mathrm{s}\over \mH}\int_0^\pi \!\!\!
       |e_{j',\alpha}(\theta_B')|^2\sin\theta_B'\,\mathrm{d}\theta_B',
%     \right],
\eea
where $\alpha=0,\pm1$,
 $e_{j,0}=e_{j,z}$ is the $z$-component of $\bm{e}_j$, and
$e_{j,\pm1}=(e_{j,x}\pm \mathrm{i} e_{j,y})/\sqrt{2}$
are the circular components.
The cross sections $\sigma_\alpha$ depend on 
$\omega$, but not on $j$ or $\theta_B$. 

The total scattering opacity from mode $j$ is
$\kappa_j^\mathrm{s}=\kappa_{j1}^\mathrm{s}+\kappa_{j2}^\mathrm{s}$,
and the total extinction opacity is
$\kappa_j=\kappa_j^\mathrm{a}+\kappa_j^\mathrm{s}$.

In the \emph{diffusion approximation} (whose accuracy was studied, e.g.,
by \citealp{SZ95}), the effective opacity
is 
\beq
   \kappa_j^{\mathrm{eff}}=
(\cos^2\vartheta/\kappa_j^\| + \sin^2\vartheta/\kappa_j^\perp)^{-1},
\eeq
where $\vartheta$ is the angle between $\bm{B}$ 
and the intensity gradient, 
\beq
 \frac {1}{\kappa_j^\|} = \frac34 \int_0^\pi
    {\cos^2\theta_B\over\kappa_j(\theta_B)}\,
    \sin\theta_B\mathrm{d}\theta_B,
\quad
  \frac{1}{\kappa_j^{\perp\vphantom{\|}}} = \frac32 \int_0^\pi
    {\sin^3\theta_B\over\kappa_j(\theta_B)}\,
    \mathrm{d}\theta_B.
\label{kappa-eff}
\eeq
The effective opacity for the nonpolarized radiation is
$\kappa^{\mathrm{eff}}=2/[(\kappa_1^{\mathrm{eff}})^{-1}+
(\kappa_2^{\mathrm{eff}})^{-1}]$.

In a partially ionized atmosphere, the opacity is contributed
by electrons, ions, and bound species. 
The scattering cross section includes contributions from the
electrons and protons: 
$
   \sigma_\alpha^\mathrm{s} = \sigma_\alpha^\mathrm{s,e}
              + \sigma_\alpha^\mathrm{s,p}$
(the Rayleigh scattering by atoms can be important only at 
 lower photon energies than considered in this paper).
The absorption cross section $\sigma_\alpha^\mathrm{a}$
includes contributions from absorption by plasma electrons and protons
(free-free transitions due to the electron-proton collisions,
 $\sigma_\alpha^\mathrm{ff}$, and proton-proton collisions,
 $\sigma_\alpha^\mathrm{pp}$),
transitions between discrete states of an atom
(bound-bound absorption, $\sigma_\alpha^\mathrm{bb}$) and
photoionization (bound-free, $\sigma_\alpha^\mathrm{bf}$).
So, for the hydrogen atmosphere, 
we can write
%% \bea&&
\beq
   \sigma_\alpha^\mathrm{a} = 
   x_\mathrm{H} (\sigma_\alpha^\mathrm{bb}+\sigma_\alpha^\mathrm{bf})
   + (1-x_\mathrm{H})\,(\sigma_\alpha^\mathrm{ff}
   + \sigma_\alpha^\mathrm{pp}) ,
\label{kappa-mix}
\eeq
where $x_\mathrm{H}$ is the number fraction of atoms,
which will be evaluated in the
following section.

% ***********************************************************
%AA% \section{Equation of state}
       \section{Equation of State}
\label{sect-EOS}
% ***********************************************************
%AA% \subsection{Calculation of tables}
      \subsection{Calculation of Tables}
Our treatment of the ionization equilibrium and EOS 
of hydrogen in strong magnetic fields is based
on the theory developed in Paper I and briefly exposed
in Sect.\ \ref{sect-thermomodel}. 
Since our free-energy model is computationally expensive,
it is not possible to use the EOS code ``on line'' in any practical
application. The alternative is to tabulate 
thermodynamic quantities covering the density, temperature, and
magnetic field domain of interest and to rely on an interpolation procedure in the table.
Here we present EOS tables
which cover a range of $\rho$, $T$, and $B$
appropriate for most typical neutron stars, such as isolated
pulsars. 

As discussed in Paper I, our model becomes less reliable 
at relatively low $T$ and high $\rho$, particularly because of formation 
of molecules and chains H$_n$, 
which are treated in an approximate manner.
In this domain, the partial number fractions and 
thermodynamic quantities are strongly model-dependent.
However, this domain of uncertainty is unimportant
for modeling of not too cold neutron stars,
because the temperature grows inside the stellar envelope.
For this reason, we have chosen as an input parameter
the ``astrophysical density parameter'' $R=\rho_0/T_6^3$,
which is customary in the stellar opacity calculations
(e.g., \citealp{Seaton,OPAL}), 
and restricted the calculation to $R<4\times10^3$.
% ***********************************************************
%AA% \subsection{Ionization equilibrium}
         \subsection{Ionization Equilibrium}
\label{sect-ioneq}
\begin{figure}\epsscale{1.}
\plotone{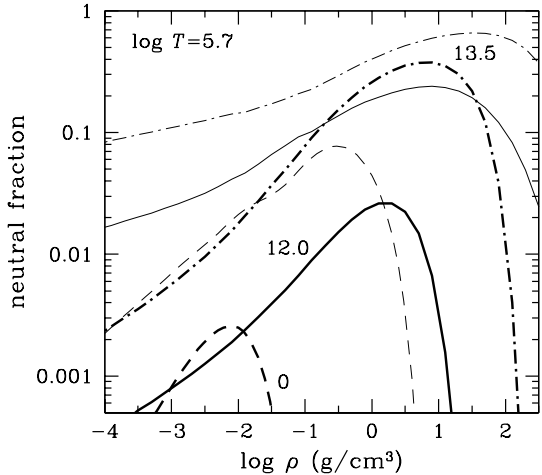}
\caption{Neutral fraction of ground-state H atoms (heavy lines)
and the total neutral fraction (including excited states and
states forming the optical pseudo-continuum; light lines) 
as function of density
at $T=5\times 10^5$ K and $B=0$ (dashed lines), $10^{12}$ G
(solid lines), and $10^{13.5}$ G (dot-dashed lines).
\label{fig-ie-b}}
\end{figure}

\begin{figure*}\epsscale{1.}
\plotone{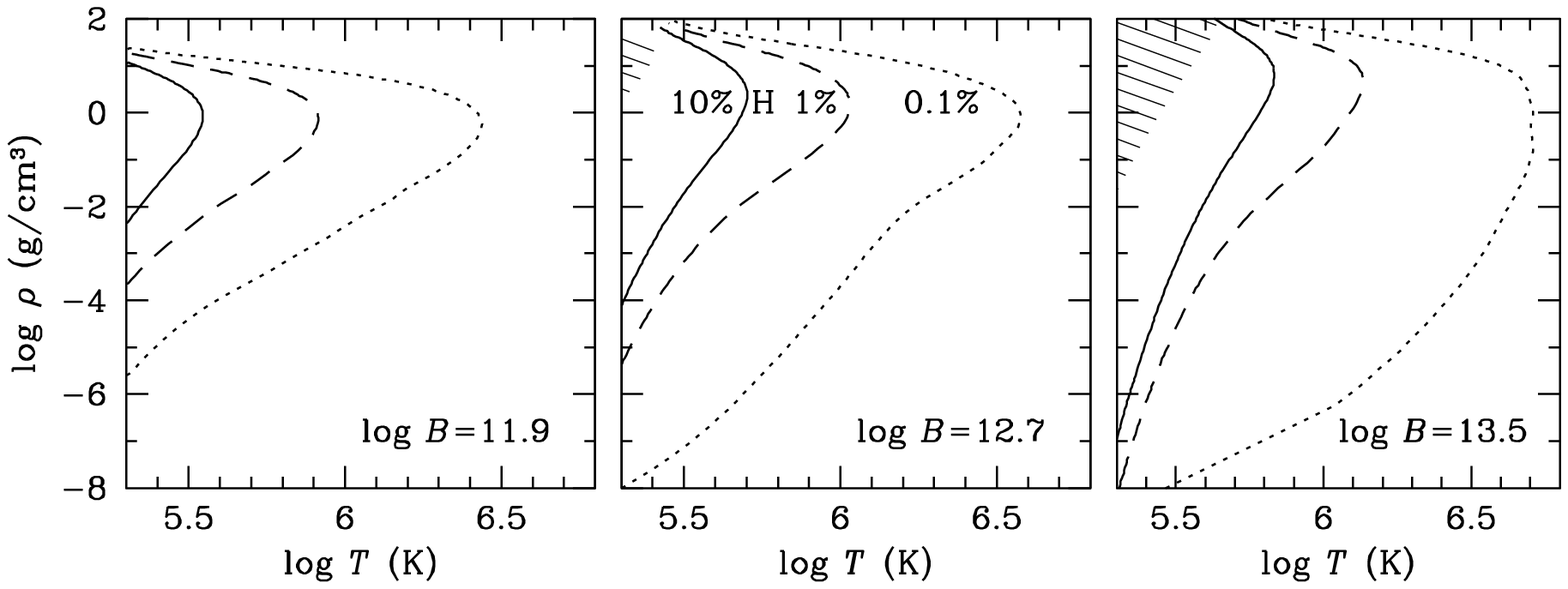}
\caption{Domains of partial ionization 
at $\log_{10} B/\mathrm{G} =11.9$, 12.7, and 13.5. 
The contours delimit the domains where the atomic fraction
exceeds 0.1\% (dotted lines), 1\% (dashed lines), or 10\% (solid lines).
Hatched is the domain where the molecular fraction exceeds 1\%.
\label{fig-phase3}}
\end{figure*}

Our detailed thermodynamic model shows that a strong magnetic field 
generally increases the fraction of bound species.
In Fig.\ \ref{fig-ie-b}, ionization equilibrium curves 
at $B=10^{12}$ G and $10^{13.5}$ G
are compared with the case of $B=0$.
The latter case is treated in the framework of
 the nonmagnetic free energy model (Sect.\ II of Paper~I),
 which is a variant of the \citet{SC91,SC92} model.
In all cases, the excited atoms contribute significantly
at low $\rho$. In the strong magnetic field,
the population of decentered atoms
is also significant at low $\rho$.
At higher density, the excluded-volume effect eliminates
the excited and decentered atoms. 
At these high densities,
 the plasma species strongly interact,
 which leads to appearance of a significant fraction of
 clusters. Such clusters contribute to the EOS
 similarly to the atoms, lowering the pressure, but their
 radiation-absorption properties are clearly different from 
 those of an isolated atom. Therefore they
 should be excluded from $x_\mathrm{H}$ in \req{kappa-mix}.
 Analogously, at low $\rho$ we should not include in $x_\mathrm{H}$
 the highly excited states that do not satisfy the 
\citet{InglisTeller} criterion of spectral line 
merging, being strongly perturbed by plasma microfields.
Such states form the so called optical pseudo-continuum 
(e.g., \citealp*{dappen87}).
This distinction between the ``thermodynamic''
and ``optical'' neutral fraction is inevitable in the
chemical picture of a plasma at high densities
(see, e.g., \citealp{P96b} for a discussion).
We discriminate the atoms which keep their identity
from the ``dissolved'' states
(i.e., strongly perturbed by the plasma environment) 
using the occupation probability formalism.
At every $s$, $\nu$, and $K_\perp$, we calculate 
the ``optical'' occupation probability
$w_{\nu s}^\mathrm{o}(K_\perp)$, replacing the Inglis--Teller criterion
by an approximate criterion based on the average atomic 
size [Eq.~(14) of \citet{PP95}].
The fraction of weakly perturbed atoms,
which contribute to the bound-bound and 
bound-free opacities, constitutes a fraction
$w_{\nu s}^\mathrm{o}(K_\perp)/w_{\nu s}^\mathrm{t}(K_\perp) < 1$ 
of the total number of atoms.
Here, $w_{\nu s}^\mathrm{t}(K_\perp)$
is the ``thermodynamic'' occupation probability derived from 
the free energy (Paper I).
 Heavy lines in Fig.\ \ref{fig-ie-b} show the 
 neutral fraction of the weakly perturbed atoms in their ground state,
 which contribute to the opacities as isolated atoms,
 whereas the light lines show the total fraction of protons
 bound in atoms or clusters.

According to our model, at relatively low $T$, pressure ionization proceeds
via a first-order phase transition.
This ``plasma phase transition'' occurs
at temperature below  
$T_\mathrm{c}\approx 3\times10^5\,B_{12}^{0.39}$ K at densities
around $\rho_\mathrm{c}\approx 143\,B_{12}^{1.18}$ \gcc\ (Paper I).
In general, the validity of the free-energy models
in the framework of the chemical picture of plasmas 
is questionable near the plasma phase transition domain.
However, the $T_\mathrm{c}$ and $\rho_\mathrm{c}$ values correspond to
$\log_{10} R_\mathrm{c}\approx 3.7$, which is beyond the upper $R$
limit for our tables,
so that the plasma phase transition
is not crossed along the tabulated isotherms.

Figure \ref{fig-phase3} shows the domains of partial ionization
in the $T$--$\rho$ plane
at three values of $B$. With increasing $B$, the domains
where the atomic fraction is above a specified level
expand significantly. For instance, at $B=5\times10^{12}$~G
the domain where $x_\mathrm{H} > 0.01$
extends to $T=10^6$ K. Such amount of atoms
 can give an important contribution to radiative opacities.

Our tables provide values of $x_\mathrm{H}$, as well as 
the fractions of ground-state atoms, molecules, and clusters
at every $R$, $T$, and $B$ entry.

% **********************************************************
%AA% \subsection{Thermodynamic functions}
       \subsection{Thermodynamic Functions}
\begin{figure*}\epsscale{1.}
\plotone{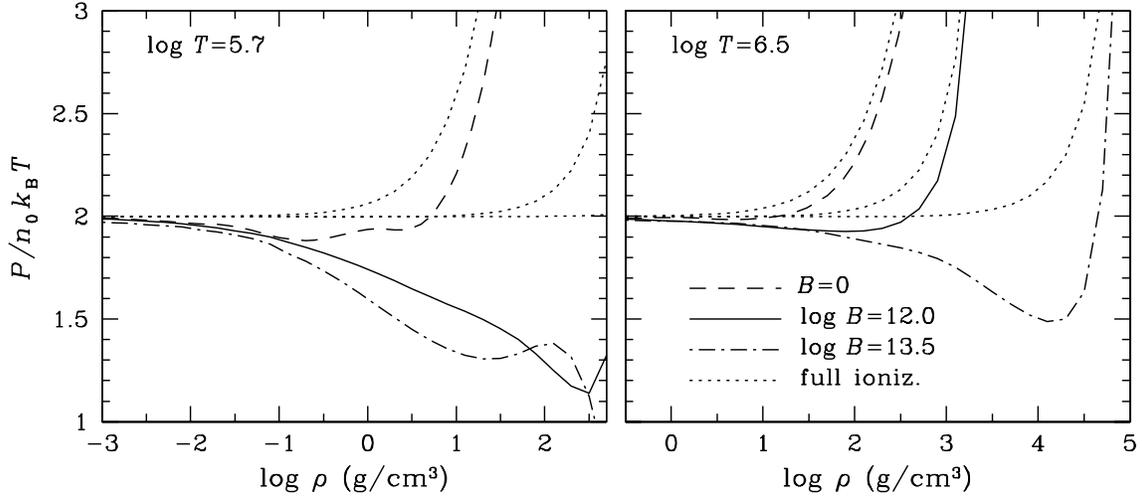}
\caption{Pressure $P$ relative to $n_0 \kB T$,
where $n_0$ is the total
number density of protons (free and bound). 
Left panel: $T=10^{5.7}$ K,
right panel: $T=10^{6.5}$ K;
dashed lines: $B=0$, solid lines: $B=10^{12}$ G,
dot-dashed lines: $B=10^{13.5}$ G.
Dotted lines represent the pressure of a fully ionized 
electron-proton ideal gas at the same values of $T$ and $B$.
\label{fig-eos-blin}}
\end{figure*}

Figure \ref{fig-eos-blin} shows pressure $P$ along two isotherms
for the same field strengths as in Fig.\ \ref{fig-ie-b}.
The pressure varies over many orders of magnitude in the
shown density range.
Therefore, in order to make 
the discussed effects more visible,
 we plot in Fig.\ \ref{fig-eos-blin}
the \emph{ratio} of $P$ to $n_0\kB T$,
the pressure of an ideal monatomic 
hydrogen gas at the same $\rho$ and $T$.

At different field strengths (including $B=0$), the deviations from 
the ideal gas behavior are qualitatively the same,
but quantitatively different. At very low $\rho$,
we have nearly fully ionized, almost ideal nondegenerate
gas of electrons
and protons, so that $P/n_0\kB T \approx2$. With increasing density,
atomic recombination proceeds according to the Saha 
equation [in the strong magnetic field, the modified Saha equation is given by
Eq.\ (54) of Paper I]. Therefore, the ratio $P/n_0\kB T$
 decreases. At high density,
however, the atoms become pressure ionized: in this region
the increase of $P$ due to the increased number of free electrons and
protons competes with a negative nonideal contribution,
which is mainly due to the Coulomb term $F_\mathrm{ex}^\mathrm{C}$ 
in the free energy. 
The dot-dashed curve bends down near the high-$\rho$ edge
of the left panel of Fig.\ \ref{fig-eos-blin} because of 
enhancement of the nonideal contribution; it is a precursor
of the plasma phase transition, where our model becomes inapplicable.
Finally, at still higher densities,
the electrons become degenerate and raise the pressure
far above $n_0\kB T$.
% (for nonrelativistic 
% degenerate electrons,
% $P\propto n_\mathrm{e}^{5/3}$ at $B=0$,
% and $P\propto n_\mathrm{e}^3$ in the strongly quantizing
% magnetic field). 
The dotted lines in Fig.\ \ref{fig-eos-blin}
show the pressure of an ideal electron-proton gas;
their upward bending marks the onset of electron degeneracy.
At $B=0$, it occurs at 
$\rho\gtrsim \mH(2\mel\kB T)^{3/2}/(3\pi^2\hbar^3)
\approx 6\,T_6^{3/2}$ \gcc,
but in the strongly quantizing field the electrons become degenerate
at $\rho\gtrsim (\mH/\pi^2\hbar\am^2)\, (\mel\kB T/2)^{1/2}
\approx 613\,B_{12}\,\sqrt{T_6}$ \gcc.

Since the strong magnetic field enhances
atomic recombination and delays pressure ionization 
 and electron degeneracy
at high densities, the discussed features of 
the $P/n_0\kB T$ curves
become more prominent and shift to higher $\rho$
with increasing $B$, as seen in Fig.\ \ref{fig-eos-blin}.

\begin{figure}\epsscale{1.}
\plotone{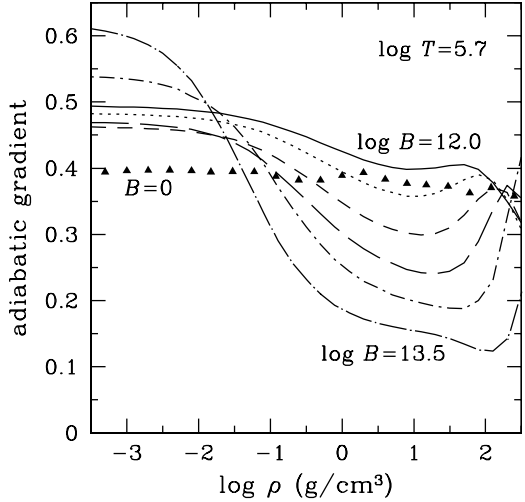}
\caption{Adiabatic temperature gradient 
$\nabla_\mathrm{ad}$ at $T=5\times10^5$ K,
for different field strengths (shown by different line styles):
$\log_{10}B/\mathrm{G} = 12.0$, 12.3, 12.6, 12.9, 13.2, and 13.5.
Triangles show the zero-field case.
\label{fig-tegrad}}
\end{figure}

Along with the pressure, our EOS tables contain internal energy $U$,
entropy $S$, specific heat $C_V$, and
the logarithmic derivatives of pressure  
$\chi_T=(\partial\ln P/\partial\ln T)_V$
and
   $\chi_\rho=-(\partial\ln P/\partial\ln V)_T$.
Other second-order quantities
can be calculated using the Maxwell relations
(e.g., \citealp{LaLi-stat1}).
For example, the heat capacity at constant pressure, $C_P$,
and adiabatic gradient, 
$\nabla_\mathrm{ad}=(\partial\log T/\partial\log P)_S$,
are given by relations
\beq
   C_P = C_V+ { PV \over T } \, {\chi_T^2\over\chi_\rho},
\quad
   \nabla_\mathrm{ad} =
        { \chi_T \over \chi_T^2 + \chi_\rho\, C_V T /(PV) }.
\label{tegrad1}
\eeq
Figure \ref{fig-tegrad} shows $\nabla_\mathrm{ad}$
at different values of $B$. At low density, 
the magnetic field increases the adiabatic gradient,
thus stabilizing the matter against convection.
This thermodynamic effect is additional to the hydromagnetic stabilization 
considered, e.g., by \citet{Chandra-hydromagnetic}
and \citet{Miralles}.
However, at higher densities, $\rho\sim (1$---100) \gcc,
there is a significant decrease of $\nabla_\mathrm{ad}$
due to the partial recombination of H atoms. 
The adiabatic gradient increases again at still higher densities,
where the plasma is fully pressure-ionized.

% ***********************************************************
%AA% \section{Cross sections}
       \section{Cross Sections}
\label{sect-cross}
% ***********************************************************
\subsection{Scattering}

The scattering cross sections by the electrons under the conditions
typical for photospheres of the neutron stars with strong magnetic
fields were thoroughly studied in the past
 (e.g., \citealp{Ventura79,Kam82,Mesz}, and references therein).
The cross section $\sigma_{-1}^\mathrm{s,e}$
exhibits a resonance
at the electron cyclotron frequency $\omc$.
Outside of a narrow (about the Doppler width) frequency range 
around $\omc$, the cross sections are
\beq
    \sigma_\alpha^\mathrm{s,e} =
          {\omega^2\over(\omega+\alpha\omc)^2
             +\nu_{\mathrm{e},\alpha}^2}\, \sigma_\mathrm{T},
\label{sigma-se}
\eeq
where $\sigma_\mathrm{T}=(8\pi/3)(e^2/\mel c^2)^2$
is the nonmagnetic Thomson cross section, and
$\nu_{\mathrm{e},\alpha}$ is an effective damping frequency
given by \req{broad} below.

The scattering by the protons, which is completely negligible
in a nonmagnetized plasma, becomes important in the strong
magnetic field, because $\sigma_{+1}^\mathrm{s,p}$
exhibits a resonance at $\omp$. It 
is the same as the electron cyclotron resonance but a different
mass and opposite charge 
of the particle:
\beq
    \sigma_\alpha^\mathrm{s,p} =
      \left( \frac{\mel}{\mpr}\right)^2
          {\omega^2\over(\omega-\alpha\omp)^2
             +\nu_{\mathrm{p},\alpha}^2}\, \sigma_\mathrm{T}.
\label{sigma-sp}
\eeq
The damping frequency $\nu_{\mathrm{p},\alpha}$ will be derived below
[\req{nu-p}].

We neglect the Doppler broadening of these resonances.
Within the thermal width of the cyclotron resonance,
the treatment of radiation scattering is nontrivial
(e.g., \citealp{Ventura-ea85},
and references therein). However,
at $T\lesssim10^7$ K, the Doppler width $\sim\omega\,\sqrt{T/T_\mathrm{r}}$
is smaller than the frequency resolution of our opacity tables
(chosen to be $\Delta\log_{10}\omega=0.02$).

% ***********************************************************
%AA% \subsection{Absorption by atoms}
        \subsection{Absorption by Atoms}
\label{sect-bb+bf}

Oscillator strengths for the H atom which rests in a strong 
magnetic field 
were calculated, e.g., by \citet{Forster}.
Bound-bound transitions of the H atom moving arbitrarily
in a strong magnetic field were studied 
by \citet{PP95}.
The modification of the binding
energies due to the atomic motion (Sect.\ \ref{sect-H-atom}) leads to 
a dramatic ``magnetic broadening'' of the spectral lines
averaged over all states of motion,
which exceeds by orders of magnitude the usual Doppler broadening.
Thus the spectral profile of the 
 bound-bound opacities becomes continuous in a wide frequency range,
 resembling a reversed bound-free profile. 
 Our calculation of the bound-bound absorption cross sections
 relies on the theory presented by \citet{PP95}
and employs fitting formulae for the binding energies,
 oscillator strengths,
 and electron collision widths derived by \citet{P98}.

Photoionization cross sections of the nonmoving H atom in 
a magnetic field were calculated by many authors (e.g.,
\citealp*{PPV},
and references therein).
Photoionization of the H atom in a strong magnetic field
with allowance for motion was studied, using different modifications  
of the adiabatic approximation (Sect.\ \ref{sect-H-atom})
by \citet{BP94}
and 
\citet*{KVH}.
A complete numerical treatment beyond the adiabatic approximation
has been developed by \citet{PP97},
who adapted the $R$-matrix formalism 
\citep{Rmatrix} 
 to the case under study.
They showed that none of the versions of the adiabatic approximation
can provide accurate $\sigma_\alpha^\mathrm{bf}$ for all
values of $\alpha$ and $K_\perp$, particularly because the
continuum-channel coupling strongly affects $\sigma_\pm^\mathrm{bf}$
at sufficiently large $K_\perp$.
Here we use the complete numerical treatment.
Since it is computationally involved,
we use an interpolation across a precalculated set of tables.
For each of the three basic polarizations, we have calculated 
$\sigma_\alpha^\mathrm{bf}(\omega,K_\perp,B)$ on a predefined grid,
with 
$\log_{10}\hbar\omega\ [\mathrm{eV}]$ ranging from 1.0 to 4.5
with step 0.02, $\log_{10}K_\perp\ [\mathrm{a.u.}]$
ranging from 1 to 3 with step 0.1,
and $\log_{10}B[\mathrm{G}]$
ranging from 11.9 to 13.5 with step 0.1.
At the low-$K_\perp$ end of the grid, $K_\perp=10$ a.u.,
the atomic properties are virtually the same as at $K=0$;
beyond the upper limit, $K_\perp>10^3$ a.u., the contribution
of the bound-free transitions to the total opacities is negligible.
The grid is sufficiently fine for calculation of atmosphere models.
However, the step 0.02 in $\log_{10}\omega$
 does not allow us to resolve the narrow Beutler--Fano-type 
resonances
which appear due to autoionizing states in the vicinity of photoionization thresholds
of partial cross sections
\citep{PPV,PP97}. 
Whenever such a resonance occurs near a grid point,
it produces a spurious outlier on the otherwise smooth
$\sigma_\alpha^\mathrm{bf}(\omega,K_\perp)$ dependence.
We filter out such outliers by smoothing $\sigma_\alpha^\mathrm{bf}$
as a function of $\omega$ at every $K_\perp$,
using the 3-point median filter.
Since the grid does not allow us to resolve
the photoionization threshold accurately, the threshold frequency 
$\omega_\mathrm{th}$
is determined independently for every
$K_\perp$, using analytic
 fits to the binding energies \citep{P98}.

In addition to the bound-bound and bound-free atomic transitions,
in a plasma environment there are transitions from bound states
to the highly perturbed atomic states 
discussed in Sect.\ \ref{sect-ioneq}.
These perturbed levels effectively dissolve and merge
in a pseudo-continuum, which lies below the photoionization threshold.
In order to take into account the radiative transitions into
this pseudo-continuum, we employ a below-threshold extrapolation, 
as described for the zero-field case
by \citet{dappen87}, \citet{Stehle}, and 
\citet{Seaton}. Below 
$\omega_\mathrm{th}$, 
the effective ``bound--quasi-free''
photoabsorption cross section
due to the dissolved lines is
\beq
   \sigma_\alpha^\mathrm{bf} (\omega < \omega_\mathrm{th})
    = \frac{2\pi e^2}{\mel c}\,
   \frac{w_i^\mathrm{o}-w_f^\mathrm{o}}{w_i^\mathrm{o}}\,
   f_{if,\alpha}\,\frac{\dd\nu_f}{\dd\omega},
\eeq
where $w_i^\mathrm{o}$ and $w_f^\mathrm{o}$ are the optical occupation 
probabilities of the initial and final states, respectively,
$f_{if,\alpha}$ is the corresponding oscillator strength,
and $\dd\nu_f/\dd\omega$ is the number of final states 
per unit frequency interval. By analogy with 
\citet{Stehle}, we interpolate $w_f^\mathrm{o}$ as function 
of frequency and set $w_f^\mathrm{o}=w_i^\mathrm{o}$
at $\omega$ smaller than the lowest allowed transition frequency.
Taking into account that $\sigma_\alpha^\mathrm{bf}(\omega<\omega_\mathrm{th})$
is a smooth continuation of $\sigma_\alpha^\mathrm{bf}(\omega>\omega_\mathrm{th})$
\citep{dappen87},
we write
\beq
    \sigma_\alpha^\mathrm{bf}(K_\perp,\omega<\omega_\mathrm{th}) = \left(
     1- \frac{w_f^\mathrm{o}(K_\perp,\omega)}{w_i^\mathrm{o}(K_\perp)} 
     \right)
    \sigma_\alpha^\mathrm{bf,extr}(K_\perp,\omega),
\label{sigma-bq}
\eeq
where $\sigma_\alpha^\mathrm{bf,extr}(K_\perp,\omega)$
is a power-law extrapolation of 
$\sigma_\alpha^\mathrm{bf}(K_\perp,\omega)$
at $\omega<\omega_\mathrm{th}$.

Unlike the $B=0$ case, in our case $f_{if,\alpha}$
and $w_f^\mathrm{o}$ depend on polarization. 
For photoabsorption by an atom in the ground state,
$f_{if,\pm1}\neq0$ only for even upper states,
whereas $f_{if,0}\neq0$ only for odd states. 
We take into consideration 
only the appropriate states while calculating $w_f^\mathrm{o}$
in \req{sigma-bq}.

The $K_\perp$-dependent cross sections are averaged over the 
distribution of atoms over $K_\perp$
with statistical weights 
$w_i^\mathrm{o}(K_\perp)\exp[-\epsilon_i(K_\perp)/\kB T]$,
as in \citet{PP95}
and \citet{PP97}.

% **********************************************************
%AA% \subsection{Free-free absorption}
       \subsection{Free-Free Absorption}
In the classical cold plasma approximation (e.g., 
\citealp{Ginzburg}), the free-free absorption by electrons is
\beq
   \sigma_\alpha^\mathrm{ff} =
          \frac{1}{(\omega+\alpha\omc)^2
             +\nu_{\mathrm{e},\alpha}^2}\, 
          \frac{4\pi e^2 
          }{ 
            \mel\, c}
            \,\nu_{\alpha}^\mathrm{ff}(\omega) ,
\label{sigma-a}
\eeq
where $\nu_{\alpha}^\mathrm{ff}$ 
is an effective frequency of electron-proton 
collisions which lead to absorption of
photons.
Broadening of the electron cyclotron resonance
in Eqs.\ (\ref{sigma-se}) and (\ref{sigma-a}) 
is determined by the sum of the effective frequencies
for absorption and scattering, 
\beq
   \nu_{\mathrm{e},\alpha}=\nu_{\alpha}^\mathrm{ff}
   +\nu_\mathrm{e}^\mathrm{s},
\label{broad}
\eeq
where 
\beq
   \nu_\mathrm{e}^\mathrm{s}=\frac{2}{3}\,\frac{e^2
   }{
         \mel c^3}\,\omega^2
\label{radfreq}
\eeq
is the natural (radiative) width of the resonance.
The damping frequency given by \req{broad}
ensures the correct value of the cyclotron absorption
cross section integrated across the resonance 
(e.g., \citealp{Ventura79}):
\beq
   \int_{\omc-\Delta\omega}^{\omc+\Delta\omega}
    [\sigma_{-1}^\mathrm{s,e} (\omega) + 
            \sigma_{-1}^\mathrm{ff}(\omega)] \,\dd \omega
         = \frac{4\pi^2e^2}{\mel\,c},
\label{cycl-osc-str}
\eeq
with $\nu_{\mathrm{e},\alpha} \ll \Delta\omega \ll \omc$.

The values of $\nu_{\alpha}^\mathrm{ff}$ are provided 
by quantum-mechanical calculations. 
It is customary 
(e.g., \citealp{Armstrong})
to express $\sigma^\mathrm{ff}$ through the
thermally averaged Gaunt factors $\bar{g}$,
or equivalently, Coulomb logarithms $\Lambda=(\pi/\sqrt{3})\,\bar{g}$.
Taking into account \req{sigma-a}, we can write
\beq
   \nu_\alpha^\mathrm{ff} =
          \frac{4}{3}\,\sqrt{\frac{2\pi}{\mel \kB T}}\,
          \frac{n_\mathrm{e}\,e^4}{\hbar \omega}\,
          \left(1-\mathrm{e}^{-u}\right)\, 
         \Lambda_\alpha^\mathrm{ff},
\quad
   u\equiv\frac{\hbar\omega}{\kB T}.
\label{nuff}
\eeq
The factor $(1-\mathrm{e}^{-u})$
allows for the induced radiation. 

% **********************************************************
%AA% \subsubsection{Infinite proton mass approximation}
        \subsubsection{Infinite Proton Mass Approximation}
In the zero-field case, the electron 
free-free absorption rate can be calculated assuming 
the electron scattering off a fixed Coulomb potential.
In this case, allowance for the finite ion mass
consists in replacing $\mel$ by the reduced mass $\mu$.
The Born
approximation yields
the well-known formula for the cross section of free-free
photoabsorption by an electron having an initial momentum $p_i$
and final momentum $p_f=(p_i^2+2\mel\hbar\omega)^{1/2}$
 (e.g., \citealp{BetheSalpeter}),
\beq
   \sigma^\mathrm{ff}(p_i,\omega) = 
   \frac{16\pi^2\,n_\mathrm{e} e^6}{3\,\mel c \hbar \omega^3}\,
  \frac{1}{p_i}
   \ln\left|\frac{p_f+p_i}{p_f-p_i}\right|,
\eeq
whose averaging over the Maxwell distribution
gives the classical Coulomb logarithm
\beq
\Lambda_\mathrm{cl}^\mathrm{ff} = \mathrm{e}^{u/2}K_0(u/2),
\label{GauntBorn}
\eeq
 $K_0(u/2)$ being the modified Bessel function.
\citet{Hummer88} calculated $\bar{g}^\mathrm{ff}$
using accurate non-Born quantum-mechanical results by
\citet{KarzasLatter} and fitted
it by a Pad\'e formula.
The nonmagnetic
Gaunt factor is applicable if
the magnetic field is nonquantizing ---
i.e., if $\beta_\mathrm{e} < 1$, where  $\beta_\mathrm{e}$
is given by \req{beta-e}.
In the quantizing magnetic fields, the Coulomb logarithm
was evaluated in the Born approximation
by several authors
\citep{PavlovPanov,Nagel,Mesz}. 
In this approximation, $\Lambda_\alpha^\mathrm{ff}$, which is generally
a function of $B$, $T$, and $\omega$, depends only
on the two dimensionless arguments, $u$ and 
$\beta_\mathrm{e}$\footnote{%
      The set of equations (\ref{magaunt}) is equivalent to 
   Eq.\ (27) of \citet{PavlovPanov}, but
   in \req{A1} for $A_n^{\pm1}$
   we have restored power 2 of $\zeta$, lacking
 in \citet{PavlovPanov}
   apparently due to a misprint.
}:
\begin{subequations}
\label{magaunt}
\beq
\Lambda_\alpha^\mathrm{ff} = \frac{3}{4}\,e^{u/2}
 \sum_{n=-\infty}^\infty 
 \int_0^\infty Q_n^\alpha(\beta_\mathrm{e},u,y)
  \,\dd y,
\eeq
where
\begin{eqnarray}
&&
 Q_n^\alpha (\beta_\mathrm{e},u,y) = \frac{y}{\zeta}\,
 \frac{A_n^\alpha }{ 
 [(y+\theta+\zeta)\,\mathrm{sinh}(\beta_\mathrm{e}/2)]^{|n|}}
\\&&
A_n^{0} = \frac{x_n K_1(x_n) }{ y+\beta_\mathrm{e}/4},
\quad
A_n^{\pm1} = \frac{y+\theta+|n|\zeta }{ \zeta^2}\,K_0(x_n),
\label{A1}
\\&&
\zeta=\sqrt{1+2\theta y+y^2},
\quad
   \theta = \frac{ 1+\exp(-\beta_\mathrm{e}) }{
       1-\exp(-\beta_\mathrm{e}) } ,
\\&&
   x_n = |u-n\beta_\mathrm{e}| \,\sqrt{0.25+y/\beta_\mathrm{e}}.
\end{eqnarray}
\end{subequations}

Equation (\ref{magaunt}) has been derived
assuming that an electron scatters off a fixed Coulomb center.
Actually the protons are moving and can absorb radiation
during collisions. Although this process is negligible at $B=0$,
it may be important
 at $\omega$ near or below $\omp$. 
The previous authors \citep{Pavlov95,Zane00,Zane01,Ozel01,HoLai}
supplemented $\sigma_\alpha^\mathrm{ff}$ by
 a cross section of ``ion free-free absorption,''
$\sigma_\alpha^\mathrm{ff,p}$. For hydrogen, taking into account 
Eqs.\ (\ref{sigma-se}) and (\ref{sigma-sp}), their formulae can be
written as
\beq
   \sigma_\alpha^\mathrm{a}=\sigma_\alpha^\mathrm{ff}
   +\sigma_\alpha^\mathrm{ff,p},
\quad
   \sigma_\alpha^{\mathrm{ff,p}} / \sigma_\alpha^\mathrm{s,p}
   = \sigma_\alpha^\mathrm{ff} / \sigma_\alpha^\mathrm{s,e} .
\label{scaling}
\eeq
We find, however, that \req{scaling} is erroneous.

% **********************************************************
%AA% \subsubsection{Absorption in proton collisions}
       \subsubsection{Absorption in Proton Collisions}
\begin{figure}\epsscale{1.}
\plotone{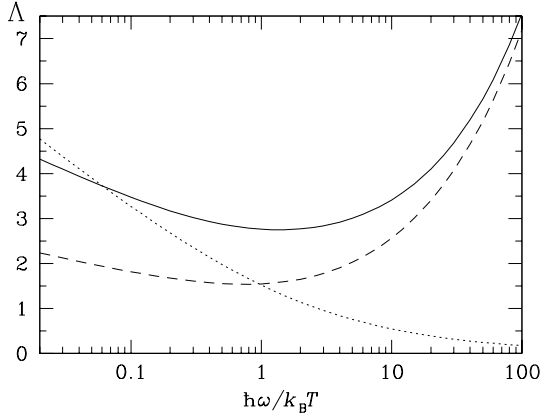}
\caption{Coulomb logarithms for photoabsorption
in collisions of nonrelativistic protons ($\Lambda_\mathrm{pp}$,
solid line),
collisions of identical but distinguishable particles (dashed line),
and electron-proton scattering 
($\Lambda_\mathrm{cl}^\mathrm{ff}$, dotted line).
\label{fig-qlpp}}
\end{figure}

There are two effects of the finite proton mass on the absorption:
first, the absorption can occur in proton-proton collisions,
and second, the absorption in the electron-proton collisions
is modified because of the proton motion. Let us start with
the first process. By analogy with \req{sigma-a}, we write
the cross section as
\beq
   \sigma_\alpha^\mathrm{pp} =
          \frac{1}{(\omega-\alpha\omp )^2
             +\nu_{\mathrm{p},\alpha}^2}\, 
          \frac{4\pi e^2
          }{ 
            {\mpr}\, c}\, \nu_{\alpha}^\mathrm{pp}(\omega) ,
\label{sigma-pp}
\eeq
where $\nu_{\alpha}^\mathrm{pp}$ is the effective frequency 
to be determined.

In the classical picture
\citep{Ginzburg}, an ion-ion 
 collision does not change
the total electric current and hence does not cause
dissipation. Therefore, it does not contribute to the damping
of radiation.
In quantum mechanics, this corresponds to vanishing
dipole matrix element for the absorption. 
In Appendix \ref{sect-pp} we evaluate the nonvanishing
quadrupole term under the condition that the magnetic field does 
not quantize proton motion, that is $\beta_\mathrm{p} < 1$, and obtain
\beq
\nu^\mathrm{pp} =  \frac{256}{3}\,\sqrt{\frac{\pi}{\mpr\kB T}}\,
   \frac{n_\mathrm{p} e^4}{\hbar\omega}\,\frac{\kB T}{\mpr c^2}\,
   (1-\mathrm{e}^{-u})\,
   \Lambda_\mathrm{pp},
\label{nupp}
\eeq
where $\Lambda_\mathrm{pp}$ is an appropriate Coulomb logarithm.
In Fig.\ \ref{fig-qlpp},
$\Lambda_\mathrm{pp}$ is plotted by the solid line.
For comparison, we also show $\Lambda$ for distinguishable
particles [corresponding to \req{sigma2}; dashed line], 
and the classical Coulomb logarithm $\Lambda_\mathrm{cl}^\mathrm{ff}$
[\req{GauntBorn}; dotted line].
The fit
\beq
   \Lambda_\mathrm{pp} \approx 0.6\,\ln(22\,u^{-1}+9\,u^{-0.3})
   +0.4\,\sqrt{\pi u}
\label{Lambda-pp}
\eeq
accurately reproduces $\Lambda_\mathrm{pp}$ at small and large $u$
and has a maximum error within 1.5\% at intermediate $u$.

We see that $\sigma_\alpha^\mathrm{pp}$
differs from $\sigma_\alpha^\mathrm{ff,p}$
[defined by \req{scaling} with use of Eqs.\ (\ref{sigma-a})
and (\ref{nuff})]
by a factor of
$32\,\sqrt{2\mpr/\mel}\,(\kB T/\mpr c^2 )\,
\Lambda_\mathrm{pp}/\Lambda_\mathrm{cl}^\mathrm{ff}
\sim T/10^9$~K.

One should remember that Eqs.\ (\ref{sigma-pp})--(\ref{Lambda-pp})
are obtained in the nonrelativistic approximation.
Therefore they do not take into account spin-flip processes
and are inapplicable
at very high $T$ or high $\omega$, where the relativistic 
corrections can be important.

% **********************************************************
%AA% \subsubsection{Electron free-free process with allowance for
%AA% finite proton mass}
       \subsubsection{Electron Free-Free Process with Allowance for
        Finite Proton Mass}
\label{sect-ff}
Let us write the cross section of photoabsorption 
due to the electron-proton collisions in the form 
of combined Eqs.\ (\ref{sigma-a}) and (\ref{nuff}), 
neglecting broadening:
\beq
   \sigma_\alpha^\mathrm{ff} =
          \frac{2^{9/2} \pi^{3/2}\,
          n_\mathrm{e}\,e^6\
          }{
          3 \mel^{3/2} (\kB T)^{1/2}\, 
             c\,\hbar \,\omega
          }\,\
          \frac{ 1-\mathrm{e}^{-u}
          }{ 
     (\omega+\alpha\omc)^2}\,
         \Lambda_\alpha^\mathrm{e}.
\label{sigma-e}
\eeq
The superscript `e' indicates that 
the electron (not proton) cyclotron resonance has been 
separated off $\Lambda_\alpha$. Since the colliding electron
and proton are treated on equal footing, we anticipate that
thus defined \emph{normalized cross section}
 $\Lambda_\alpha^\mathrm{e}$ will reveal
a resonant peak at $\omega\sim\omp$.

\begin{figure}\epsscale{1.}
\plotone{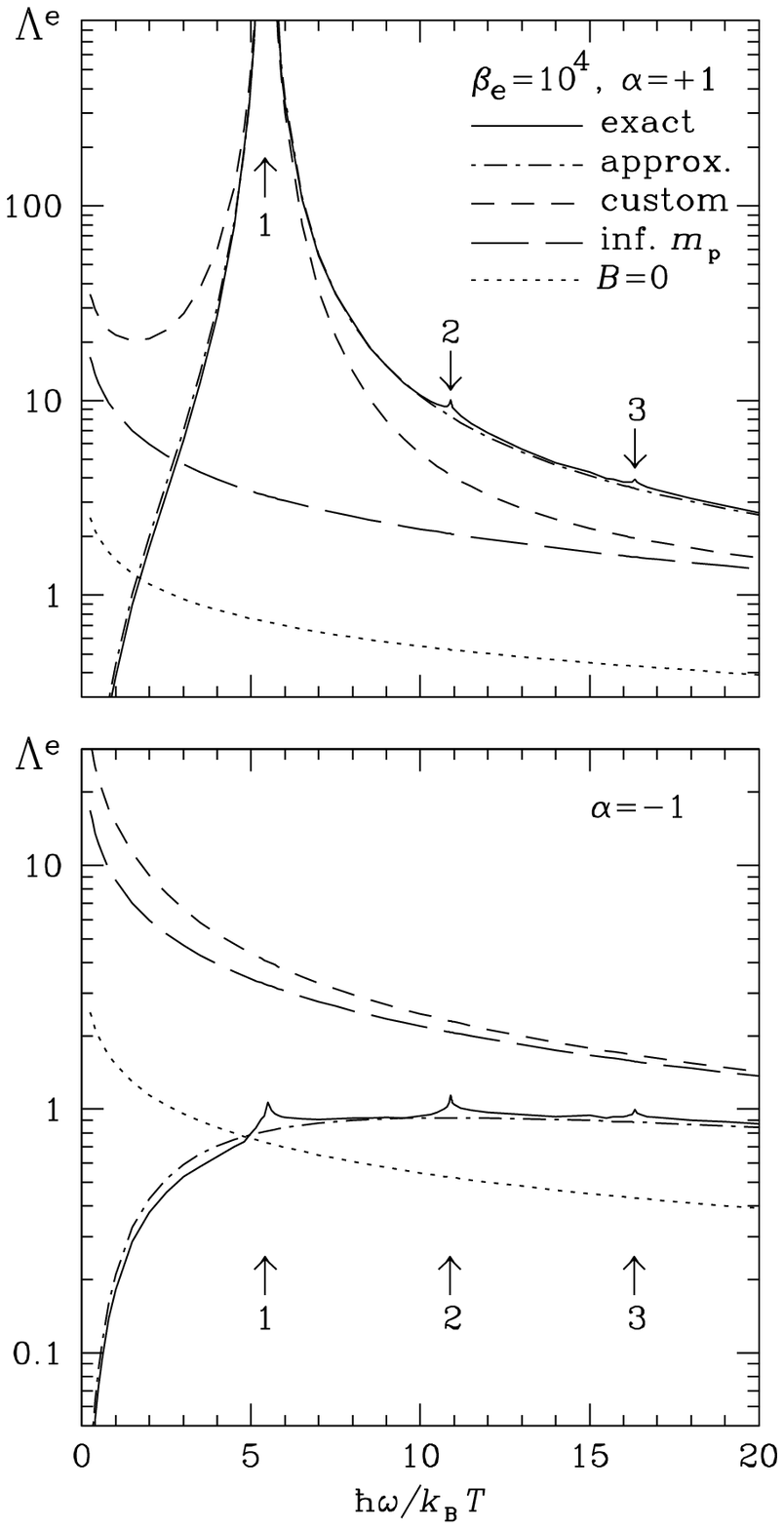}
\caption{Comparison of accurate and 
approximate normalized free-free
cross sections $\Lambda_\alpha^\mathrm{e}$
[\protect\req{sigma-e}]
for two circular polarizations $\alpha=\pm1$
at $\beta_\mathrm{e}=10^4$.
Solid line: Eqs.\ (\protect\ref{sigma-int})--(\protect\ref{w});
dot-dashed line: \protect\req{sigma-fit};
short-dashed line: customary approximation, \req{scaling};
long-dashed line: $\Lambda_\alpha^\mathrm{ff}$, \protect\req{magaunt};
dotted line: $\Lambda_\mathrm{cl}^\mathrm{ff}$,
\protect\req{GauntBorn}. Arrows indicate 
the proton cyclotron harmonics.
\label{fig-ql02}}
\end{figure}

The initial and final states of the
interacting electron and proton are just continuum states
of the H atom. 
An accurate treatment of these states would imply
a solution of the coupled-channel equations (\ref{system})
and calculation of the $R$-matrix, as we did
for the bound-free process.
However, we will restrict to the first Born approximation.
In this approximation, $\Lambda_\alpha^\mathrm{e}$
is given by Eqs.\ (\ref{Lambda-e})--(\ref{w})
derived in Appendix \ref{sect-ep}.
For the longitudinal polarization ($\alpha=0$),
a calculation by these equations well
reproduces the Coulomb logarithm (\ref{magaunt})
obtained in the infinite proton mass approximation.
However, for the two circular polarizations the result is different;
it is shown in Fig.\ \ref{fig-ql02} by the solid lines.
As expected, we see a remarkable proton cyclotron resonance
at $\alpha=+1$, which is due to the denominator $\omega-\omp$
in the last terms of \req{w+}. In addition, 
for both circular polarizations there are smaller spikes
at higher proton-cyclotron harmonics, arising from 
the logarithmic singularities of $\tilde{v}_{nsn's}(\rho,\varkappa)$
at $\varkappa\to0$ [see \req{vss}]. Apart from these spikes,
$\Lambda_{\pm1}^\mathrm{e}$ is accurately described
by the formula
$\Lambda_{\pm1}^\mathrm{e}\approx 
\Lambda_{\pm1}^\mathrm{ff}\,
\omega^2/(\omega-\alpha\omp)^2$, where $\Lambda_{\pm1}^\mathrm{ff}$
is given by \req{magaunt} (dot-dashed lines in the figure).
We note that the factor $\omega^2/(\omega-\alpha\omp)^2$ 
naturally appears in the classical plasma model with allowance
for the ion motion. The classical model
also helps to restore the damping factors neglected in \req{sigma-e}.
Let $\nu_\mathrm{e}$ and $\nu_\mathrm{p}$ be the electron
and proton damping frequencies due to processes other than
the electron-ion collisions. For the processes
considered above (Thomson scattering and proton-proton collisions%
\footnote{%
We do not consider the electron-electron collisions.
At $B=0$, they are known to be
unimportant for the bremsstrahlung, except for
relativistic energies (e.g., \citealp{BetheSalpeter}).
The magnetic field does not change this conclusion,
because the resonance, that appears at $\omega=\omc$ for this process,
merges in the more powerful classical cyclotron resonance
for the usual free-free and Thomson processes.}),
we have $\nu_\mathrm{e}=\nu_\mathrm{e}^\mathrm{s}$
and $\nu_\mathrm{p}=\nu_\mathrm{p}^\mathrm{s}+\nu^\mathrm{pp}$,
where $\nu_\mathrm{p}^\mathrm{s}=2e^2 \,\omega^2 / (3\,\mpr \,c^3)$
is the natural width of the proton cyclotron resonance.
Averaged Newtonian equations of motion for the electrons
and protons in the magnetic and radiation fields
give the complex permittivity tensor $\varepsilon$ 
(cf.\ \citealp{Ginzburg}, \S10).
Neglecting $\sqrt{\mel/\mpr}$ compared to unity and
assuming that $\nu_\mathrm{p}\ll\nu_\mathrm{e}\ll \omega$,
we have
\bea
&&\hspace*{-1em}
   \mathrm{Im}\,(\varepsilon_{xx}+\mathrm{i}\alpha\varepsilon_{xy})
%\nonumber\\&&
      = \frac{\omega_\mathrm{pl}^2}{\omega}\,
      \Big[\,\omega_\mathrm{cp}\,(\omc+\alpha\omega)\,\nu_\mathrm{p}
     + \omega\,(\omega-\alpha\omega_\mathrm{cp})\,\nu_\mathrm{e} 
\nonumber\\&&
   + \omega^2\,\nu_\alpha^\mathrm{ff} \,\Big] 
\,\Big/\,
     \Big\{\big[\,(\omega + \alpha\omc)\,(\omega-\alpha\omc) 
 \nonumber\\&&\hspace*{5em}
         - \nu_\alpha^\mathrm{ff}\,(\nu_\mathrm{p}+\nu_\mathrm{e}\,\mel/\mpr)
       - \nu_\mathrm{p}\,\nu_\mathrm{e}
      \,\big]^2
\nonumber\\&&
     +\big[\,\omega\,\nu_\alpha^\mathrm{ff} 
     + (\omega+\alpha\omc)\,\nu_\mathrm{p}
     + (\omega-\alpha\omega_\mathrm{cp})\,\nu_\mathrm{e}\,\big]^2\Big\} ,
\eea
where $\alpha=\pm1$, and Im means the imaginary part. 
Neglecting the tiny shift of the position of the cyclotron 
resonances caused by the damping, we can now write
\beq
   \sigma_\alpha^\mathrm{ff} \approx
          \frac{\omega^2
          }{
          (\omega+\alpha\omc)^2\,(\omega-\alpha\omp)^2
             +\omega^2 \tilde\nu_\alpha^2}\, 
          \frac{4\pi e^2 \nu_{\alpha}^\mathrm{ff}
          }{ 
            \mel\, c},
\label{sigma-fit}
\eeq
where 
\beq
   \tilde\nu_\alpha = \nu_{\alpha}^\mathrm{ff}
        + (1+\alpha\omc/\omega)\,\nu_\mathrm{p}
        + (1-\alpha\omp/\omega)\,\nu_\mathrm{e},
\eeq
and
$\nu_{\alpha}^\mathrm{ff}$ is given by \req{nuff}.

The accurate calculations according to Appendix \ref{sect-ep}
are reproduced by Eqs.\ (\ref{sigma-fit}) and (\ref{nuff}), if to multiply 
$ \Lambda_\alpha^\mathrm{ff}$ in the latter equation by a correction factor
of the order of unity. This quantum correction factor
 proves to be the same for $\alpha=+1$ and $\alpha=-1$. Thus, the 
two effective collision frequencies (longitudinal $\nu_0^\mathrm{ff}$ 
and transverse $\nu_{+1}^\mathrm{ff}=\nu_{-1}^\mathrm{ff}$)
provide the three $\sigma_\alpha^\mathrm{ff}$.

Near the electron cyclotron resonance ($\omega\approx\omc$,
$\alpha=-1$), the damping frequency $\tilde\nu_\alpha$ approximately reproduces 
$\nu_{\mathrm{e},\alpha}$ in \req{broad}, which ensures the condition
(\ref{cycl-osc-str}). Near the proton cyclotron resonance ($\omega\approx\omp$,
$\alpha=+1$), the effective damping frequency is
$\tilde\nu_\alpha \omega/\omc\approx \tilde\nu_\alpha\mel/\mpr
\approx\nu_{\mathrm{p},\alpha}$,
where
\beq
   \nu_{\mathrm{p},\alpha} = \nu_\alpha^\mathrm{ff,p}
   + \nu_\mathrm{p}^\mathrm{s} + \nu_\alpha^\mathrm{pp},
\quad
%  \nu_\alpha^\mathrm{ff,p} = (\mel/\mpr)\,\nu_\alpha^\mathrm{ff},
  \nu_\alpha^\mathrm{ff,p} \equiv
   \frac{\mel}{\mpr}\,\nu_\alpha^\mathrm{ff}.
\label{nu-p}
\eeq
Equation (\ref{nu-p}) is consistent with
the requirement of oscillator strength conservation for
the proton cyclotron resonance in polarization $\alpha=+1$,
fully analogous to \req{cycl-osc-str}. 

From the relations
\beq
   \frac{\nu^\mathrm{pp}}{\nu_\alpha^\mathrm{ff,p}}
    = 1.78\times10^{-4}\,T_6 
    \,\frac{\Lambda_\mathrm{pp}}{\Lambda_\alpha^\mathrm{ff}},
\quad
   \frac{\nu^\mathrm{pp}}{\nu_\mathrm{p}^\mathrm{s}} = 
   \frac{3.6\rho_0}{T_6^{5/2}}\,\frac{1-\mathrm{e}^{-u}}{u^3}
   \,\Lambda_\mathrm{pp},
\eeq
we see that the proton-proton collisions can be safely neglected
at any $T$ and $\rho$ typical of outer envelopes of the neutron stars.

\begin{figure}\epsscale{1.}
\plotone{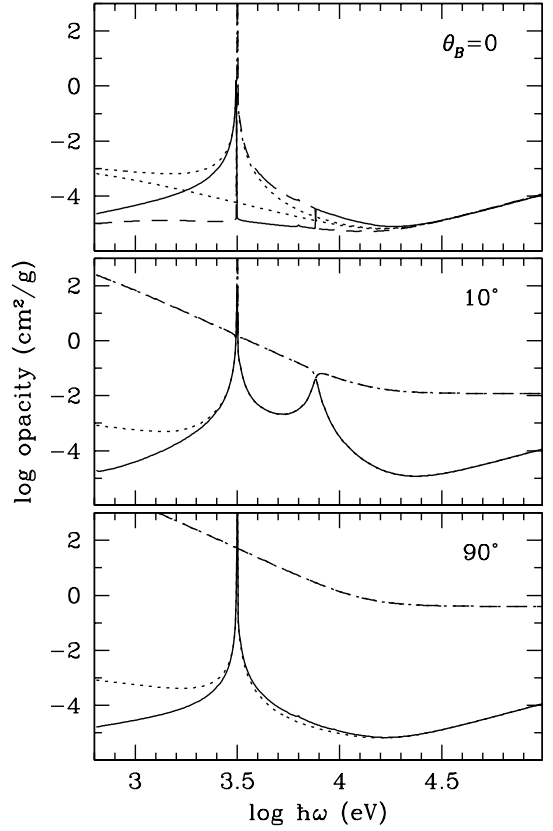}
\caption{Monochromatic radiative opacities
for the ordinary (dashed lines) and extraordinary (solid lines)
polarization modes in a plasma at $\rho=500$ \gcc,
$T=5\times10^6$ K, and $B=5\times10^{14}$ G,
for $\theta_B=0$, $10^\circ$, and $90^\circ$.
Dotted lines show the opacities according to \protect\req{scaling}.
Here the opacity is calculated assuming complete ionization.
\label{fig-ang14_7}}
\end{figure}

\begin{figure*}\epsscale{.7}
\plotone{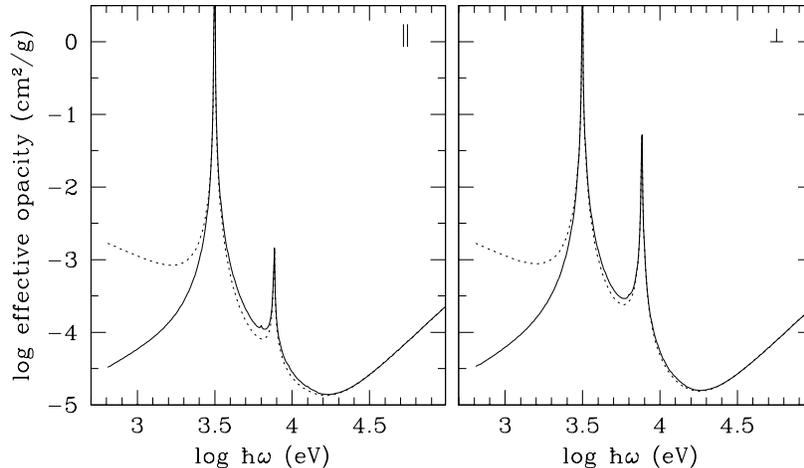}
\caption{Effective opacities for diffusion
of nonpolarized radiation along (left panel) or across (right panel)
magnetic field, at the same plasma parameters as in 
Fig.\ \protect\ref{fig-ang14_7}:
new (solid lines) and old (dotted lines) results.
\label{fig-opm14_7}}
\end{figure*}

For comparison, Fig.\ \ref{fig-ql02} also shows the  
nonmagnetic Coulomb logarithm $\Lambda_\mathrm{cl}^\mathrm{ff}$ 
(dotted line), $\Lambda_\alpha^\mathrm{ff}$ 
given by \req{magaunt} (long dashes), 
which neglects the finite proton mass,
and $\Lambda_\alpha^\mathrm{e}$ which would correspond to the
calculation of $\sigma_\alpha^\mathrm{a}$ according to 
the traditional \req{scaling} (short dashes).
It is easy to see the difference of Eqs.\ (\ref{sigma-e})
and (\ref{sigma-fit}) from \req{scaling}.
The customary recipe (\ref{scaling}) misses the interference
of the first two terms 
in each of Eqs.\ (\ref{w-}), (\ref{w+})
(related to transitions with changing Landau number $n$)
with the terms in square brackets
(related to transitions which keep $n$ constant).
At $\omega < \omp$ the latter terms tend to compensate
the former ones. 
A suppression factor $\sim(\omega/\omp)^2$, which follows
from \req{sigma-fit} at $\omega\ll\omp$,
is brought about by this interference.
From the classical physics point of view, it may be explained
as follows: very slow ($\omega\ll\omp$)
oscillations of the radiation electric field, perpendicular to the
constant magnetic field, make both particles, electron
and proton, to drift adiabatically
with the velocity $(c/B^2)\bm{E}\times\bm{B}$, 
so that in the drifting frame of reference
 they do not ``feel'' the electric field
 of the electromagnetic wave.
The suppression of the free-free
cross sections takes place for both circular polarizations.

As well known, the Born approximation is accurate
only at $\hbar\omega$ much larger than the binding energies.
In order to partly correct $\Lambda_\alpha^\mathrm{ff}$
beyond this approximation and to
recover the non-Born Gaunt factor $\bar{g}^\mathrm{ff}$ at $B=0$,
we multiply 
$\Lambda_\alpha^\mathrm{ff}$ by the ratio of 
$\bar{g}^\mathrm{ff}$ \citep{Hummer88}
to $\bar{g}_\mathrm{Born}^\mathrm{ff}
=(\sqrt{3}/\pi)\,\Lambda_\mathrm{cl}^\mathrm{ff}$.

% **********************************************************
\section{Opacities}
\label{sect-opac}
%
%AA% \subsection{Fully ionized atmosphere}
        \subsection{Fully Ionized Atmosphere}
For a fully ionized atmosphere model, the monochromatic opacities 
calculated in the framework of the theory outlined in 
Sect.\ \ref{sect-opac-basic} do not differ much
from the opacities used in the previous models (e.g., 
\citealp{Shib92}) at $\omega > \omp$.
However, the improved treatment of the free-free contribution
results in a considerable modification of the opacities
at $\omega<\omp$. Figure \ref{fig-ang14_7} shows monochromatic
opacities for two polarization modes, $j=1$ (solid lines)
and $j=2$ (dashed lines) 
for radiation propagating at three angles $\theta_B$
with respect to the field lines. 
The atmosphere parameters chosen for this figure can be
expected near the bottom of a magnetar photosphere: 
$\rho=500$ \gcc, $T=5\times10^6$ K, and $B=5\times10^{14}$ G.
An extension of our EOS for the superstrong magnetic fields  
(work currently in progress --- \citealp*{CDP02}) indicates that 
the model of a fully ionized atmosphere is adequate
for this relatively high value of $\rho$.
At these parameters, the proton cyclotron resonance 
at $\hbar\omega=3.15$ keV is quite prominent.
At small $\theta_B$, the opacities of the two normal modes
cross each other at $\omega\approx\omp$, which is a well known
phenomenon (e.g., \citealp{Shib92}).
Another mode crossing, which occurs at
$\hbar\omega\approx7.6$ keV, is due to the vacuum resonance
(e.g., \citealp{Soffel,PavlovGnedin,Shib92}).
Near the crossing points the modes
may become completely nonorthogonal
(``collapse'') at certain angles, 
so that their designation is ambiguous 
\citep{Soffel}. 
%% Here we do not consider the possible 
%% resonant mode conversion (\citealp{HoLai02}),
%% which at present remains a controversial issue
%% (\citealp{Ozel02}; Lai \& Ho \citealp{LaiHo02b}).

The dotted curves in Fig.\ \ref{fig-ang14_7} are obtained
using \req{scaling}. We see that this traditional 
calculation strongly overestimates $\kappa_1$ at large $\theta_B$,
and overestimates both $\kappa_1$ and $\kappa_2$ at $\theta_B=0$,
if $\hbar \omega$ is small enough.

Figure \ref{fig-opm14_7} shows the effective opacities $\kappa^\mathrm{eff}$
for the diffusion of nonpolarized 
radiation along ($\|$) and across ($\perp$)
magnetic field (Sect.\ \ref{sect-opac-basic}),
for the same plasma parameters as in Fig.\ \ref{fig-ang14_7}.
The peak at $\log\hbar\omega/\mbox{eV}=3.5$ 
is due to the ion cyclotron resonance, and the one
at $\log\hbar\omega/\mbox{eV}=3.9$ due to the vacuum resonance.
The barely visible intermediate spike at 
$\log\hbar\omega/\mbox{eV}=3.8$
is the quantum resonance of the Coulomb logarithm
at the doubled proton-cyclotron frequency (cf.\ Fig.\ \ref{fig-ql02}).

We see that the improvement of the free-free cross section
discussed in Sect.\ \ref{sect-ff} is important 
for the effective opacities shown in Fig.\ \ref{fig-opm14_7}
at $\omega<\omp$. At $\omega\lesssim0.3\,\omp$, the difference
exceeds one order of magnitude.
Moreover, it has an impact on the Rosseland mean opacities,
as discussed below.

% **********************************************************
%AA% \subsection{Partially ionized atmosphere}
        \subsection{Partially Ionized Atmosphere}
\begin{figure*}\epsscale{1.}
\plotone{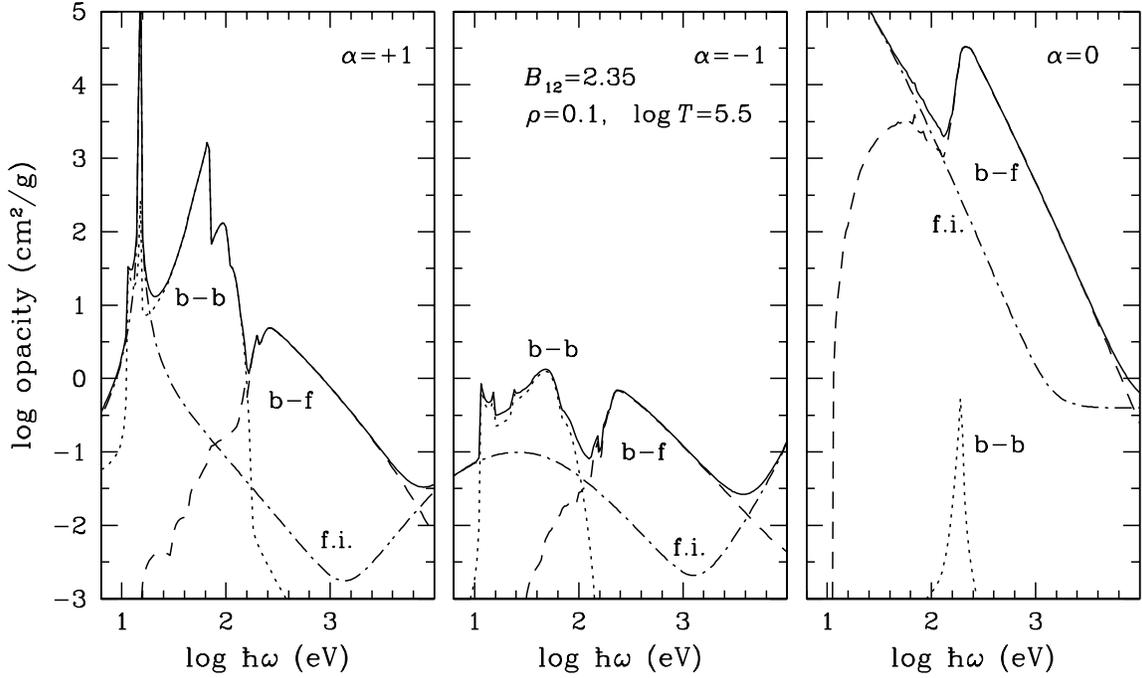}
\caption{Monochromatic opacities for the basic polarizations
($\alpha=+1$, left panel; $\alpha=-1$, middle panel;
$\alpha=0$, right panel) in a typical 
partially ionized neutron-star atmosphere
at $\rho=0.1$ \gcc, $T=10^{5.5}$ K, and $B=2.35\times10^{12}$ G.
Dot-dashed lines: opacities of fully ionized component;
dashed lines: bound-free opacities;
dotted lines: bound-bound opacities; solid lines: total opacities.
\label{fig-f12_37}}
\end{figure*}

\begin{figure*}\epsscale{1.}
\plotone{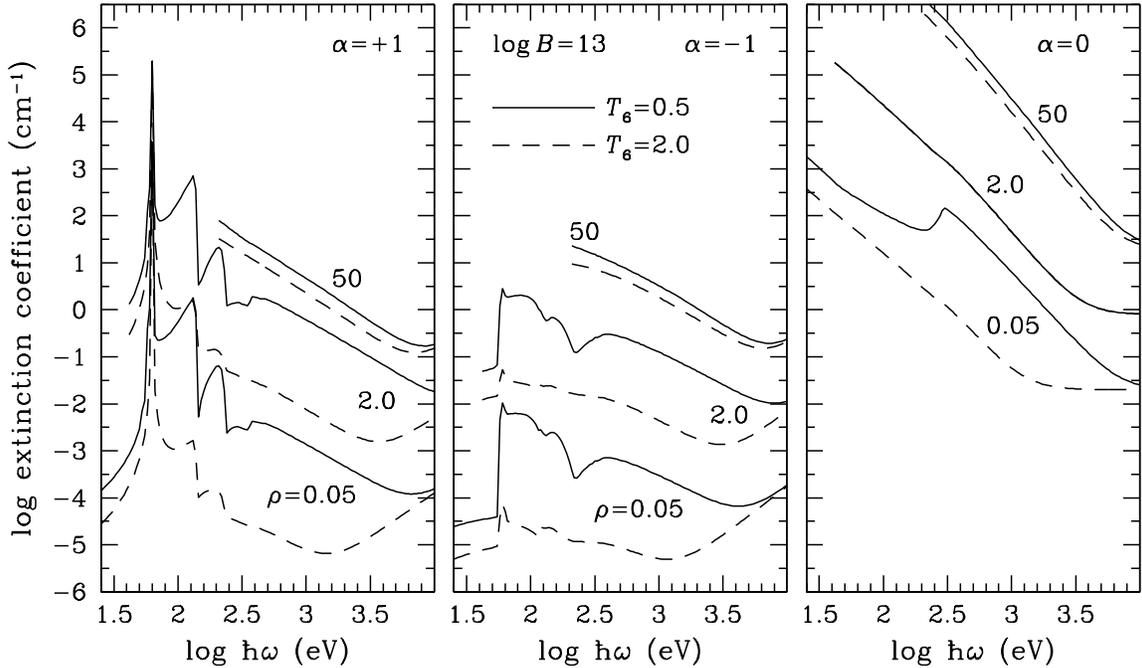} % [45 280 560 590]
\caption{Monochromatic opacities for the same basic polarizations
as in Fig.\ \protect\ref{fig-f12_37}
at $T=5\times10^5$ K
(solid lines)
and $2\times10^6$ K (dashed lines),
 $\rho=0.05$ \gcc, 2 \gcc, and 50 \gcc, and $B=10^{13}$ G.
\label{fig-f13_0}}
\end{figure*}

\begin{figure*}\epsscale{1.}
\plotone{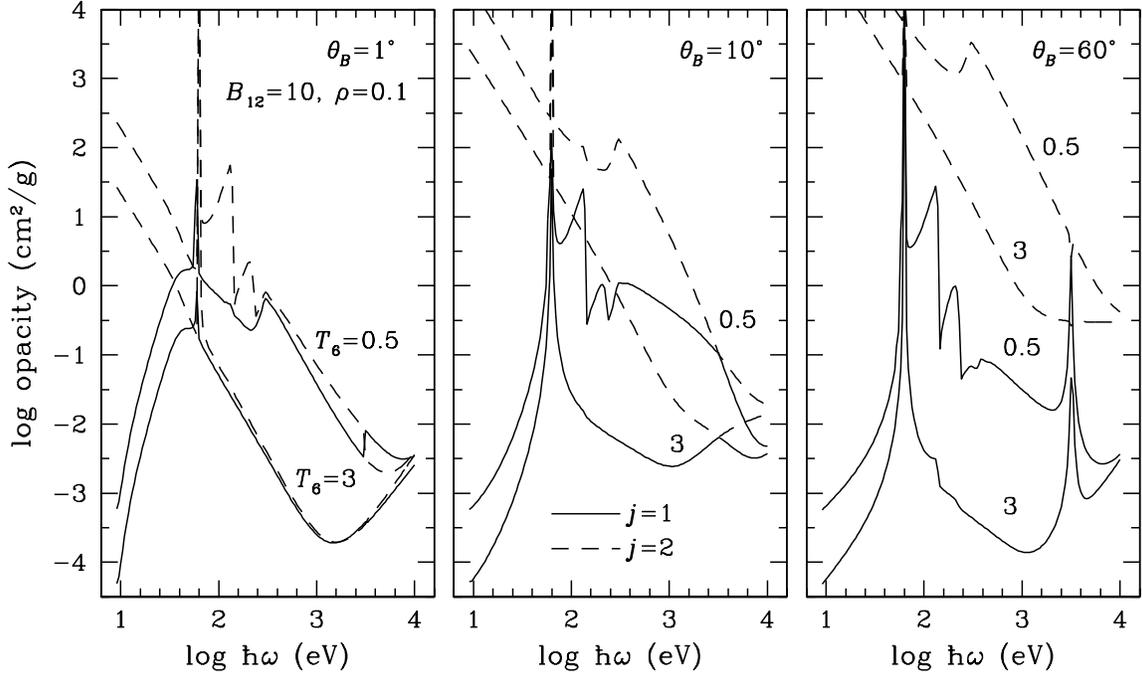} % [45 280 560 590]
\caption{Monochromatic opacities for 
the extraordinary ($j=1$, solid lines)
and ordinary ($j=2$, dashed lines) modes 
at $T=5\times10^5$ K (upper curves) and $3\times10^6$ K
(lower curves), for $\rho=0.1$ \gcc, $B=10^{13}$ G,
and $\theta_B=1^\circ$, $10^\circ$, and $60^\circ$.
\label{fig-ang13_0}}
\end{figure*}

\begin{figure*}\epsscale{.9}
\plotone{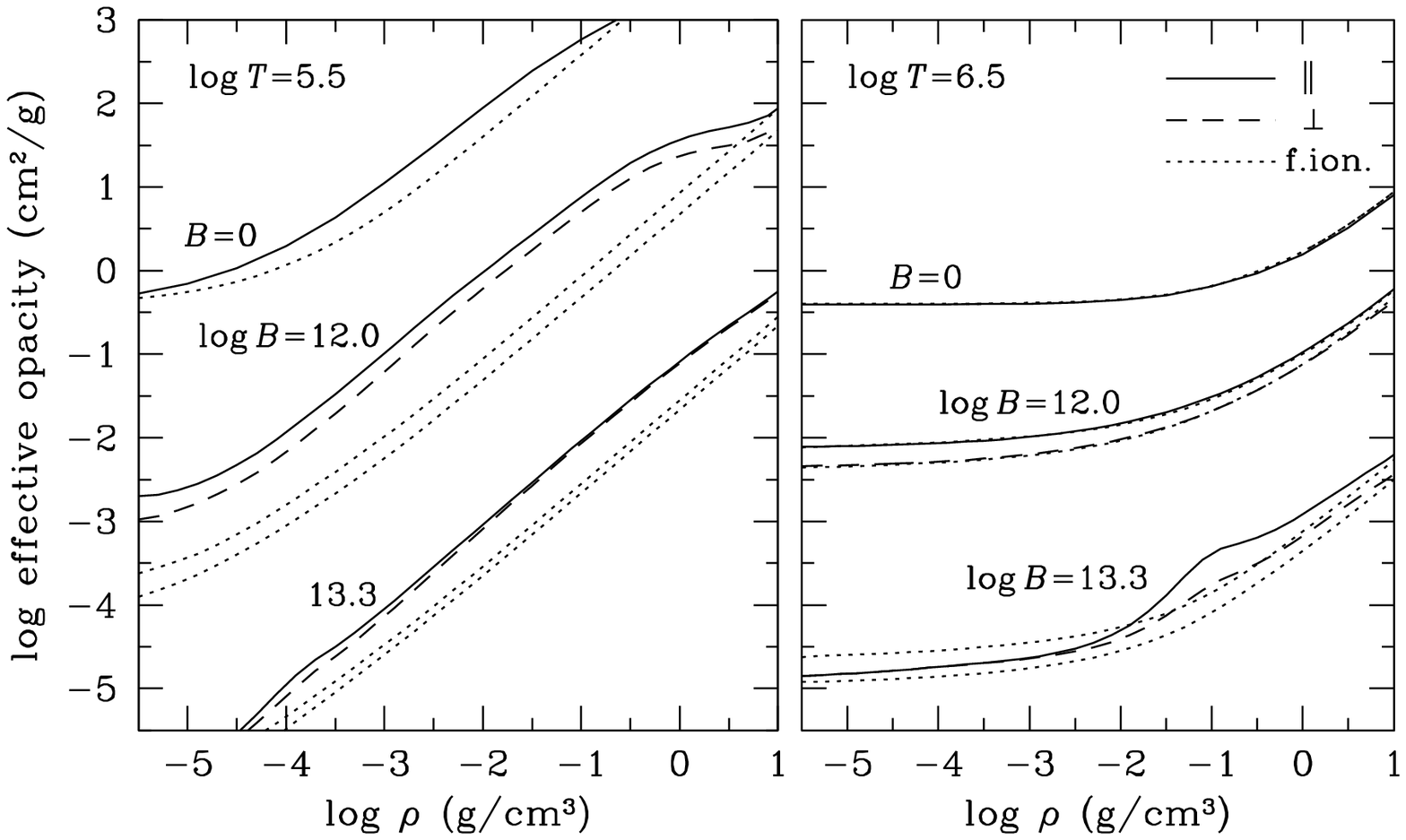} % [40 285 570 590]
\caption{Effective Rosseland mean opacities for diffusion
of nonpolarized radiation along (solid lines) or across (dashed lines)
magnetic field, at $T=10^{5.5}$ K (left panel)
and $10^{6.5}$ K, for $B=0$, $10^{12}$ G, and $2\times10^{13}$ G.
Dotted lines show the opacities of a fully ionized plasma
(at each value of $B\neq0$, the lower dotted curve corresponds
to the transverse diffusion and the upper to the longitudinal one).
\label{fig-ross}}
\end{figure*}

As follows from Sect.\ \ref{sect-ioneq}, the amount
of neutral hydrogen in neutron-star photospheres can be 
significant at $T\lesssim10^6$. For example,
at $\rho=0.1$ \gcc, $T=10^{5.5}$ K, and $B=2.35\times10^{12}$ G,
12\% of protons are bound in the ground-state H atoms.
Monochromatic opacities for this case are shown 
in Fig.\ \ref{fig-f12_37} for three basic polarizations.
The contribution of the fully ionized plasma component
is shown by dot-dashed lines, while dotted and dashed lines 
show bound-bound and bound-free contributions, respectively.
The total opacities are plotted by the solid lines.
We see that the bound-free contribution is important
for any polarization, whereas the bound-bound opacity is 
important for $\alpha=\pm1$ but unimportant for
the longitudinal polarization ($\alpha=0$).
This is because the dipole selection rule forbids
radiative transitions between different tightly bound 
states with absorption of a photon polarized along $\bm{B}$.
Transitions to the odd hydrogenlike states ($\nu=1,3,\ldots$) are
allowed, but the corresponding occupation probabilities are small,
so that these upper levels are effectively merged into the continuum.
The absorption peak at $\log_{10}\hbar\omega/\mbox{eV}\approx 1.84$
on the left panel
corresponds to the transition from the ground state
to the state with $s=1$ for the centered atoms.
It would be a narrow spectral line without atomic motion.
On the middle panel ($\alpha=-1$), 
there would be no significant bound-bound absorption 
at all, were the motion neglected.
However, the thermal motion of atoms across the field drastically
modifies the spectrum. There appears significant absorption 
for $\alpha=-1$. The bump at $\hbar\omega\approx 100$ eV 
for $\alpha=+1$ is due to 
 the transition to the second excited level ($s=2$), 
 which would be negligible for nonmoving atoms.
The magnetic broadening mentioned in Sect.\ \ref{sect-bb+bf}
smears the photoionization edges at $\hbar\omega\gtrsim100$ eV and
extends the bound-bound absorption frequency range 
down to $\sim10$ eV for
any polarization.
The spikes near this low $\omega$ are explained by the 
 $K_\perp$-dependence of the transition energy
\citep{PP95}.
 The spike at $\omega\approx\omp$ 
 ($\log_{10}\hbar\omega/\mbox{eV}\approx 1.17$)
 appears because
 the transitions between the
 decentered states whose quantum numbers $s$ differ by one
 ($\Delta s=1$) correspond to the energies 
 $\approx\hbar\omp$, almost independent of $K_\perp$
 [it follows from \req{chi}, since $E_{s\nu}^\|(K_\perp)$
 is small at large $K_\perp$].
 Another resonance occurs at a frequency
 slightly below $\omp$, which corresponds to the minimum 
 distance between the states with $\Delta s=1$ (the avoided
 crossings, cf.\ \citealp{P94}), because 
 the energy difference has zero derivative with respect to $K$ there.
 These resonances are smoothed by the electron impact
 broadening.
 
In general, we see that 
partial ionization must be taken into account
in the opacity calculations at these plasma parameters.

Figure \ref{fig-f13_0} shows the total extinction coefficients
$\rho\kappa_\alpha$
for different $\rho$ and $T$ at $B=10^{13}$ G.
The curves are truncated from the left at $\omega = \omega_\mathrm{pl}$.
At $\rho = 50$ \gcc, the opacities
are smooth functions of $\omega$, which reflects the fact 
that virtually all excited atomic levels 
are merged into continuum at these $\rho$ and $B$.
At $\rho\leq2$ \gcc, however, the curves clearly reveal
the features due to the bound-bound and bound-free
transitions.

The opacities for the three basic polarizations,
combined with appropriate components of the polarization 
vectors, provide the opacities in the two normal modes
(Sect.\ \ref{sect-opac-basic}). For example,
Fig.\ \ref{fig-ang13_0} shows the normal-mode opacities 
for $\rho=0.1$ \gcc\ and $B=10^{13}$ G,
at three values of $\theta_B$ and two values of $T$.
At the lower $T=5\times10^5$ K, the features 
arising from the bound-bound and bound-free transitions
are clearly visible at any $\theta_B$.

%AA% THE TABLE WAS HERE \begin{table}

\begin{deluxetable*}{rrrrrrrrrrrrrr}
\tabletypesize{\small}
 \tablecaption{Sample EOS and opacity table.
 \label{table}}
\tablehead{ % \multicolumn{1}{c}{$R$}
&\multicolumn{7}{c}{thermodynamic functions}
 & \multicolumn{4}{c}{number fractions}
 & \multicolumn{2}{c}{log(opacities)}\\
\multicolumn{1}{c}{(1)}&\multicolumn{1}{c}{(2)}&\multicolumn{1}{c}{(3)}&\multicolumn{1}{c}{(4)}&\multicolumn{1}{c}{(5)}&\multicolumn{1}{c}{(6)}&\multicolumn{1}{c}{(7)}&\multicolumn{1}{c}{(8)}&\multicolumn{1}{c}{(9)}&\multicolumn{1}{c}{(10)}&\multicolumn{1}{c}{(11)}&\multicolumn{1}{c}{(12)}
&\multicolumn{1}{c}{(13)}&\multicolumn{1}{c}{(14)}
}
\startdata
%    \multicolumn{2}{c}{log($T$)}  &   \multicolumn{2}{c}{log($B$)} \\
    \multicolumn{2}{c}{6.000}  &  \multicolumn{2}{c}{13.000} \\
$-7.40 $&$\!\!\!\! 0.8173 $&$\!\!\!\!1.999 $&$\!\!\!\!1.256$&$\!\!\!\!59.737$&$\!\!\!\!2.378 $&$\!\!\!\!1.001$&$\!\!\!\!1.000 $&$\!\! 3.77$E$-04$&$\!\!\!\!7.15$E$-05$&$\!\!\!\!0.00$E$+00$&$\!\!\!\!7.81$E$-03 $&$\!\!\!\!-4.896$&$\!\!\!\!-4.896$\\
$-7.20 $&$\!\!\!\! 1.0173 $&$\!\!\!\!1.999 $&$\!\!\!\!1.256$&$\!\!\!\!58.818$&$\!\!\!\!2.378 $&$\!\!\!\!1.001$&$\!\!\!\!1.000 $&$\!\! 4.15$E$-04$&$\!\!\!\!8.30$E$-05$&$\!\!\!\!0.00$E$+00$&$\!\!\!\!8.85$E$-03 $&$\!\!\!\!-4.872$&$\!\!\!\!-4.872$\\
\multicolumn{1}{c}{\ldots}&\multicolumn{1}{c}{\ldots}&\multicolumn{1}{c}{\ldots}&\multicolumn{1}{c}{\ldots}&\multicolumn{1}{c}{\ldots}&\multicolumn{1}{c}{\ldots}&\multicolumn{1}{c}{\ldots}&
\multicolumn{1}{c}{\ldots}&\multicolumn{1}{c}{\ldots}&\multicolumn{1}{c}{\ldots}&\multicolumn{1}{c}{\ldots}&\multicolumn{1}{c}{\ldots}    &\multicolumn{1}{c}{\ldots}&\multicolumn{1}{c}{\ldots}\\
$-0.20 $&$\!\!\!\! 7.9984 $&$\!\!\!\!1.914 $&$\!\!\!\!1.028$&$\!\!\!\!26.662$&$\!\!\!\!2.473 $&$\!\!\!\!1.062$&$\!\!\!\!0.984 $&$\!\! 1.19$E$-02$&$\!\!\!\!8.10$E$-03$&$\!\!\!\!6.31$E$-11$&$\!\!\!\!9.91$E$-02 $&$\!\!\!\!-1.242$&$\!\!\!\!-1.447$\\
$ 0.00 $&$\!\!\!\! 8.1949 $&$\!\!\!\!1.898 $&$\!\!\!\!0.977$&$\!\!\!\!25.736$&$\!\!\!\!2.501 $&$\!\!\!\!1.074$&$\!\!\!\!0.980 $&$\!\! 1.33$E$-02$&$\!\!\!\!9.09$E$-03$&$\!\!\!\!1.77$E$-10$&$\!\!\!\!1.06$E$-01 $&$\!\!\!\!-1.091$&$\!\!\!\!-1.295$\\
$ 0.20 $&$\!\!\!\! 8.3905 $&$\!\!\!\!1.879 $&$\!\!\!\!0.915$&$\!\!\!\!24.807$&$\!\!\!\!2.536 $&$\!\!\!\!1.088$&$\!\!\!\!0.976 $&$\!\! 1.45$E$-02$&$\!\!\!\!1.01$E$-02$&$\!\!\!\!4.26$E$-10$&$\!\!\!\!1.13$E$-01 $&$\!\!\!\!-0.954$&$\!\!\!\!-1.158$\\
$ 0.40 $&$\!\!\!\! 8.5853 $&$\!\!\!\!1.857 $&$\!\!\!\!0.841$&$\!\!\!\!23.873$&$\!\!\!\!2.575 $&$\!\!\!\!1.105$&$\!\!\!\!0.971 $&$\!\! 1.51$E$-02$&$\!\!\!\!1.08$E$-02$&$\!\!\!\!8.15$E$-10$&$\!\!\!\!1.21$E$-01 $&$\!\!\!\!-0.835$&$\!\!\!\!-1.039$\\
$ 0.60 $&$\!\!\!\! 8.7793 $&$\!\!\!\!1.831 $&$\!\!\!\!0.755$&$\!\!\!\!22.935$&$\!\!\!\!2.616 $&$\!\!\!\!1.123$&$\!\!\!\!0.966 $&$\!\! 1.45$E$-02$&$\!\!\!\!1.09$E$-02$&$\!\!\!\!1.09$E$-09$&$\!\!\!\!1.32$E$-01 $&$\!\!\!\!-0.738$&$\!\!\!\!-0.941$\\
$ 0.80 $&$\!\!\!\! 8.9721 $&$\!\!\!\!1.801 $&$\!\!\!\!0.654$&$\!\!\!\!22.000$&$\!\!\!\!2.658 $&$\!\!\!\!1.142$&$\!\!\!\!0.963 $&$\!\! 1.23$E$-02$&$\!\!\!\!9.97$E$-03$&$\!\!\!\!8.36$E$-10$&$\!\!\!\!1.44$E$-01 $&$\!\!\!\!-0.662$&$\!\!\!\!-0.866$\\
%$ 1.00 $&$\!\!\!\! 9.1646 $&$\!\!\!\!1.770 $&$\!\!\!\!0.538$&$\!\!\!\!21.063$&$\!\!\!\!2.699 $&$\!\!\!\!1.161$&$\!\!\!\!0.961 $&$\!\! 8.92$E$-03$&$\!\!\!\!7.75$E$-03$&$\!\!\!\!2.76$E$-10$&$\!\!\!\!1.55$E$-01 $&$\!\!\!\!-0.597$&$\!\!\!\!-0.802$\\
\multicolumn{1}{c}{\ldots}&\multicolumn{1}{c}{\ldots}&\multicolumn{1}{c}{\ldots}&\multicolumn{1}{c}{\ldots}&\multicolumn{1}{c}{\ldots}&\multicolumn{1}{c}{\ldots}&\multicolumn{1}{c}{\ldots}&
\multicolumn{1}{c}{\ldots}&\multicolumn{1}{c}{\ldots}&\multicolumn{1}{c}{\ldots}&\multicolumn{1}{c}{\ldots}&\multicolumn{1}{c}{\ldots}    &\multicolumn{1}{c}{\ldots}&\multicolumn{1}{c}{\ldots}\\
$ 3.00 $&$\!\!\!\!10.9665 $&$\!\!\!\!1.122 $&$\!\!\!\!-1.917$&$\!\!\!\!11.401$&$\!\!\!\!2.765 $&$\!\!\!\!1.935$&$\!\!\!\!0.692 $&$\!\! 0.00$E$+00$&$\!\!\!\!0.00$E$+00$&$\!\!\!\!0.00$E$+00$&$\!\!\!\!1.50$E$-02 $&$\!\!\!\! 2.207$&$\!\!\!\! 2.039$\\
$ 3.20 $&$\!\!\!\!11.1018 $&$\!\!\!\!0.967 $&$\!\!\!\!-2.457$&$\!\!\!\!10.364$&$\!\!\!\!2.836 $&$\!\!\!\!2.276$&$\!\!\!\!0.670 $&$\!\! 0.00$E$+00$&$\!\!\!\!0.00$E$+00$&$\!\!\!\!0.00$E$+00$&$\!\!\!\!3.76$E$-03 $&$\!\!\!\! 3.083$&$\!\!\!\! 2.904$\\
$ 3.40 $&$\!\!\!\!11.2407 $&$\!\!\!\!0.840 $&$\!\!\!\!-3.068$&$\!\!\!\! 9.334$&$\!\!\!\!2.900 $&$\!\!\!\!2.651$&$\!\!\!\!0.734 $&$\!\! 0.00$E$+00$&$\!\!\!\!0.00$E$+00$&$\!\!\!\!0.00$E$+00$&$\!\!\!\!4.11$E$-04 $&$\!\!\!\! 4.248$&$\!\!\!\! 4.055$\\
$ 3.60 $&$\!\!\!\!11.4056 $&$\!\!\!\!0.775 $&$\!\!\!\!-3.736$&$\!\!\!\! 8.301$&$\!\!\!\!2.946 $&$\!\!\!\!2.906$&$\!\!\!\!0.975 $&$\!\! 0.00$E$+00$&$\!\!\!\!0.00$E$+00$&$\!\!\!\!0.00$E$+00$&$\!\!\!\!1.40$E$-05 $&$\!\!\!\! 5.782$&$\!\!\!\! 5.574$\\
 \enddata
 \end{deluxetable*}

% **********************************************************
%AA% \subsection{Rosseland mean opacities}
      \subsection{Rosseland Mean Opacities}
Along with the thermodynamic functions
and number fractions of species, our tables contain 
Rosseland mean effective opacities for longitudinal
($\kappa_\mathrm{R}^\|$) or transverse ($\kappa_\mathrm{R}^\perp$) 
propagation of nonpolarized radiation. They are calculated
in a standard way (e.g., \citealp{Armstrong})
from the effective monochromatic opacities 
for the diffusion approximation,
$\kappa^\|$ and $\kappa^\perp$, defined in Sect.\ \ref{sect-opac-basic}.

The improvement of the free-free cross section  
substantially affects $\kappa_\mathrm{R}^\|$
and $\kappa_\mathrm{R}^\perp$.
For example, in the case of fully ionized plasma shown in 
Fig.\ \ref{fig-opm14_7}, 
we obtain $\kappa_\mathrm{R}^\|=1.8\times10^{-4}$ cm$^2$ g$^{-1}$
and $\kappa_\mathrm{R}^\perp=1.6\times10^{-4}$ cm$^2$ g$^{-1}$,
in reasonable agreement with the analytic fit 
in \citet{PY01} 
($2.1\times10^{-4}$ cm$^2$ g$^{-1}$ and
 $1.7\times10^{-4}$ cm$^2$ g$^{-1}$, respectively).
Since the latter fit did not take into account the ion cyclotron 
resonance, we conclude that this resonance is unimportant 
for the Rosseland opacities in the given example. 
Meanwhile, the traditional treatment
of this resonance [\req{scaling}] yields effective 
opacities shown by the dotted lines in Fig.\ \ref{fig-opm14_7},
whose Rosseland means are 
$\kappa_\mathrm{R}^\|=1.3\times10^{-3}$ cm$^2$ g$^{-1}$
and $\kappa_\mathrm{R}^\perp=1.0\times10^{-3}$ cm$^2$ g$^{-1}$,
--- that is, about six times larger than the accurate values.

Figure \ref{fig-ross} illustrates the density-dependence
of the Rosseland opacities at two values of $T$
for $B=10^{12}$ G and $2\times10^{13}$ G.
For comparison, the nonmagnetic OPAL opacities
(\citealp{OPAL};
\url{http://www-phys.llnl.gov/Research/OPAL/}) are shown.\footnote{%
As mentioned in Sect.\ IID of Paper I,
our model of partially ionized hydrogen plasma
at $B=0$ accurately reproduces the OPAL opacities.}
Dotted lines represent the effective Rosseland opacities
in the model of a fully ionized electron-proton plasma,
according to the fit of \citet{PY01}.

As well known, the strong magnetic field makes the atmosphere 
more transparent at given $\rho$ and $T$,
because of the presence of large $\omc$ in denominators of 
Eqs.\ (\ref{sigma-se}) and (\ref{sigma-e}).
At sufficiently large $\rho$, there is a good agreement between
the opacity tables and the analytic fully-ionized plasma model.
However, at $\rho\lesssim(1$---10) \gcc, there are large differences,
which reflect the contribution of bound-bound and/or bound-free
transitions in the effective opacities. Remarkably,
this difference is significant at $T=10^{5.5}$ K
even in the nonmagnetic case (the upper curve on the left panel).
As noted in Paper I, 
the contribution of bound species increases with increasing $B$.

In the case of $T=10^{6.5}$ K (right panel of Fig.\ \ref{fig-ross}),
the model of fully ionized plasma is quite accurate at $B\leq10^{12}$ G.
However, this is not the case
at the higher field strength $B=2\times10^{13}$ G, where the contribution
from bound species is again appreciable.

% ************************************************************
%AA% \section{Description of the tables}
       \section{Description of the Tables}
\label{sect-tables}
The input parameters for our tables are $B$, $T$,
and the astrophysical density parameter $R=\rho_0/T_6^3$.
At present, the tables are calculated for 
$11.9\leq\log_{10} B/\mathrm{G}  \leq 13.5$ 
with step $\Delta\log_{10} B = 0.1$,
$5.3 \leq \log_{10} T/\mathrm{K} \leq 7.0$ 
with step $\Delta\log_{10} T = 0.05$,
and $-7.4\leq \log_{10} R \leq 3.6$
with step $\Delta\log_{10} R = 0.2$.

The tables for different values of $T$ and $B$
 have identical structure. 
An example is shown in Table \ref{table}.
 The first line contains $\log_{10} T/\mbox{K}$ and $\log_{10} B/\mbox{G}$.
 Each row then provides:
 \begin{enumerate}
\setlength{\itemsep}{0pt}
\item $\log_{10} R$;
\item $\log_{10} P$, where $P$ is the pressure in bar$=10^6$ dyn cm$^{-2}$;
\item the dimensionless pressure parameter $PV/(N_0\kB T)$, where $N_0$ is the total number of protons
(free and bound) in volume $V$;
\item the dimensionless internal-energy parameter $U/(N_0\kB T)$;
\item the dimensionless entropy parameter $S/(N_0\kB)$;
\item the reduced heat capacity $C_V/(N_0\kB)$;
\item the logarithmic pressure derivative $\chi_T=(\partial\ln P/\partial\ln T)_V$;
\item the logarithmic pressure derivative  $\chi_\rho=(\partial\ln P/\partial\ln \rho)_T$;
\item the atomic fraction  $x_\mathrm{H}$, that is 
    the total number of H atoms with non-dissolved energy levels,
    divided by $N_0$;             
\item the ground-state atomic fraction;
\item the molecular fraction (the number of H$_2$ molecules with non-destroyed levels, divided by $N_0$);
\item the fraction of protons comprised in clusters and in strongly perturbed atoms and molecules;
\item $\log_{10}\kappa^\|$, where $\kappa^\|$ is the effective Rosseland mean opacity 
 for transport of nonpolarized radiation
 along magnetic field lines in the diffusion approximation, in cm$^2$ g$^{-1}$;
\item $\log_{10}\kappa^\perp$, where $\kappa^\perp$ is analogous 
  to $\kappa^\|$, but for diffusion of radiation 
  in the direction perpendicular to $\bm{B}$.
\end{enumerate}

We have also written a Fortran program for the cubic-polynomial 
interpolation of the tabulated data in the 3-parameter space
of $B$, $T$, and $\rho$.
The tables and the program are available at 
\url{http://www.ioffe.rssi.ru/astro/NSG/Hmagnet/}.

% ************************************************************
\section{Conclusions}
% ************************************************************
We have calculated the EOS and radiative opacities of 
fully and partially ionized hydrogen plasmas in a wide range
of densities, temperatures, and magnetic fields
typical for photospheres of the strongly magnetized neutron stars.
The first- and second-order thermodynamic functions, 
non-ionized fractions,
and effective Rosseland mean opacities
are published in the electronic form.

The opacities are calculated more accurately than in the previous
publications. In particular, we take into account  
suppression of the free-free absorption below the proton
cyclotron frequency, which was overlooked previously.
This effect reduces the opacities of the ionized component of
the plasma by orders of magnitude at photon energies
$\hbar\omega\lesssim0.3\,\hbar\omp\sim 0.02\,B_{12}$ keV,
which necessitates a revision of the previously published
models of X-ray spectra of magnetars 
\citep{Zane01,HoLai,HoLai02,Ozel01,Ozel02}.
On the other hand, the bound-bound and bound-free absorption,
neglected in the previous models of neutron-star atmospheres,
increase the opacities by more than one order of magnitude   
at $\hbar\omega\sim(0.1$--3) keV in the outer atmosphere layers of 
the ordinary neutron stars
with $B\sim10^{12}$--$10^{13.5}$ G and $T < (1$--$3)\times10^6$ K,
which can also significantly affect the spectra.

One can expect that the effect of the bound species
on the EOS and opacities is as important 
for magnetars (despite their supposedly higher temperatures) 
as for the ordinary neutron stars.
To check this, we need to extend our model to higher $B$;
preliminary high-$B$ results \citep{CDP02} support this anticipation.

      \acknowledgements
We thank
Dong Lai, G\'erard Massacrier, Yura Shibanov, and Dima Yakovlev
for useful discussions, and also Dong Lai and Yura Shibanov 
for careful reading of the manuscript and valuable remarks. 
A.P.\ gratefully acknowledges hospitality
of the theoretical astrophysics group 
at the Ecole Normale Sup\'erieure de Lyon
and the Astronomy Department of Cornell University.
The work of A.P.\ is supported in part by RFBR grants 
02-02-17668 and 00-07-90183.
%AA% \end{acknowledgements}

%\appendix
\begin{appendix}

\section{A. Photoabsorption due to Proton-Proton Collisions}
\label{sect-pp}
The general formula for the differential cross section
of absorption of radiation by a quantum-mechanical system
 is (e.g., \citealp{Armstrong})
\beq
   \dd\sigma=\frac{4\pi^2}{\omega c}
   \left|\bm{e}\cdot\langle f| \bm{j}_\mathrm{eff} | i \rangle
   \right|^2\,\delta(E_f-E_i-\hbar\omega)\,\dd\nu_f,
\label{dsigma}
\eeq
where $i$ and $f$ are the initial and final states of 
the system, $\dd\nu_f$ is the density of final states,
$\bm{e}$ is the polarization vector, 
$\bm{j}_\mathrm{eff}=
\mathrm{e}^{\mathrm{i}\bm{k}_\gamma\cdot\bm{r}} \bm{j}$,
$\bm{j}$ is the electric current operator, and $\bm{k}_\gamma$
is the photon wave number.
For two charged particles in a magnetic field,
\beq
   \bm{j}_\mathrm{eff}=
   (\mathrm{e}^{\mathrm{i}\bm{k}_\gamma\cdot\bm{r}_1}
   q_1\,\bm{\pi}_1/m_1 + 
   \mathrm{e}^{\mathrm{i}\bm{k}_\gamma\cdot\bm{r}_2}
   q_2\,\bm{\pi}_2/m_2),
\eeq
where $q_i$ and $m_i$ are the particle charge and mass ($i=1,2$),
and $\bm{\pi}_i$ is given by \req{pi_i}.
Introducing, in the standard way, the center-of-mass 
($\bm{R}$, $\bm{P}$)
and relative ($\bm{r}$, $\bm{p}$) coordinates and momenta
of two protons and using the gauge 
$\bm{A}(\bm{r}_i)=(1/2)\,\bm{B}\times\bm{r}_i$, we get
\bea
   \bm{j}_\mathrm{eff} &=& 
   \mathrm{e}^{\mathrm{i}\bm{k}_\gamma\cdot\bm{R}} \frac{e}{\mpr}\bigg[
  \bm{P} - \frac{e}{c} \bm{B}\times\bm{R} 
%AA% \nonumber\\&&
    + \mathrm{i}(\bm{k}_\gamma\cdot\bm{r})
    \left( \bm{p} - \frac{e}{4 c}\,\bm{B}\times\bm{r}\right)
    \bigg] + O(\bm{k}_\gamma\cdot\bm{r})^2
\eea
The first two terms do not contribute to the free-free absorption,
because they do not contain the relative variables and, therefore,
are decoupled from the Coulomb interaction.
The remaining terms are similar to $\bm{j}_\mathrm{eff}$
relevant to absorption of radiation by a particle with charge
$e/2$ and mass $\mpr$, except for the factor 
$\bm{k}_\gamma\cdot\bm{r}$, which is small
at $\hbar\omega\ll\mpr c^2$. 
The absorption cross section can be written in the form of \req{sigma-pp}.
The nonquantizing magnetic field,
$\beta_\mathrm{p} \ll1$,
does not affect the effective collision frequency
 $\nu_{\alpha}^\mathrm{pp}$, which in this case
does not depend on $\alpha$. 
We have evaluated the proton free-free cross section
in the nonrelativistic Born approximation,
using the technique of Fourier transforms 
(\citealp{BetheSalpeter}, \S77).
For two distinguishable particles of equal 
mass $\mpr$ and charge $e$, the cross section is
\bea&&
   \sigma_\ast^\mathrm{pp}(p_i,\omega) = \frac{256\pi^2}{3}\,
   \frac{n_\mathrm{p} e^6}{\mpr c \,\hbar\,\omega^3}\,
%AA% \nonumber\\&&\qquad\times
   \frac{1}{p_i}\left\{ \frac{p_i p_f}{(\mpr c)^2}
   + \frac{3}{10}\,\frac{p_i^2+p_f^2}{(\mpr c)^2}\,
   \ln\left|\frac{p_f+p_i}{p_f-p_i}\right| \right\}.
\label{sigma2}
\eea
Taking into account that the colliding protons are identical
and have the spin 1/2,
one should calculate the matrix element in \req{dsigma}
for symmetric and antisymmetric final states
and sum up the cross sections with the statistical weights
1/4 and 3/4, respectively (e.g., 
\citealp{LaLi-QM}). This leads to the equation
\beq
 \sigma^\mathrm{pp}(p_i,\omega) = 2\sigma_\ast^\mathrm{pp}(p_i,\omega)
       - \sigma_\times^\mathrm{pp}(p_i,\omega),
\label{sigma-pp-tot}
\eeq
where
\bea
&&
   \sigma_\times^\mathrm{pp}(p_i,\omega) = \frac{128\pi^2}{3}\,
   \frac{n_\mathrm{p} e^6}{\mpr c \,\hbar\,\omega^3}\,
% \nonumber\\&&\qquad\times
   \frac{1}{p_i}\bigg\{ 
     \frac65\,\frac{p_i p_f}{(\mpr c)^2}\,
     \frac{p_i^4+p_f^4}{(p_i^2+p_f^2)^2}
%E% \nonumber\\&&\qquad
     + \left(\frac{\hbar\omega}{c}\right)^2
     \frac{p_i^4+p_f^4+0.8\,p_i^2 p_f^2}{(p_i^2 + p_f^2)^3}\,
     \ln\left|\frac{p_f+p_i}{p_f-p_i}\right| \bigg\}.
\label{sigma-x}
\eea
In Eqs.\ (\ref{sigma2}) and (\ref{sigma-x}), $p_f^2=p_i^2+\mpr\hbar\omega$, 
since the reduced mass equals $\mpr/2$. 
The Maxwell distribution for the relative momenta is
$\mathcal{F}_\mathrm{pp}(p_i)=(4/\sqrt{\pi})\,(\mpr \kB T)^{-3/2}\,
p_i^2\exp(-p_i^2/\mpr\kB T)$. Averaging of \req{sigma-pp-tot}
with this distribution gives
the cross section in the form (\ref{sigma-pp}) with
$\nu^\mathrm{pp}$ given by \req{nupp},
where $\Lambda_\mathrm{pp}$ is 
calculated by averaging the $p_i$-dependent factors of
Eqs.\ (\ref{sigma2}) and (\ref{sigma-x}). 

       \section{B. Photoabsorption due to Electron-Proton Collisions}
\label{sect-ep}
In the case of photoabsorption due to 
the electron-proton collisions, the initial and final 
states in \req{dsigma}
are the continuum states of the hydrogen atom
described by \req{Schr}. 
The wave function of the relative electron-proton motion 
can be written as
\beq
   \psi(\bm{r}) = \psi_0(\bm{r}) + \psi_1(\bm{r}),
\quad
   \psi_0(\bm{r}) = \frac{\mathrm{e}^{\mathrm{i}k_0 z}}{\sqrt{L}}\,
      \Phi_{ns}(\bm{r}_\perp),
\eeq
where $\psi_0(\bm{r})$ describes free motion with $z$-component
of the relative momentum $\hbar k_0$,
$L$ is the normalization length, and $\psi_1(\bm{r})$
is a perturbation due to the Coulomb interaction.
Let us apply one-dimensional Fourier transformation
\beq
   \tilde\psi(\bm{r}_\perp,k) = \frac{1}{\sqrt{L}}
   \int_{-L/2}^{L/2} \mathrm{e}^{-\mathrm{i}kz}\,\psi(\bm{r})\,\dd z.
\eeq
In the limit of $L\to\infty$, we will have 
$\tilde\psi_0(\bm{r}_\perp,k) \to (2\pi/L)\,\delta(k-k_0)\,
\Phi_{ns}(\bm{r}_\perp)$. 
Let us use expansion (\ref{expansion}) for $\psi_1(\bm{r})$;
then
\beq
   \tilde\psi_1(\bm{r}_\perp,k) = \sum_{n's'} \tilde{g}_{n's'}(k)\,
   \Phi_{ns}(\bm{r}_\perp).
\eeq
This is equivalent to replacing $g_{n's'}(z)$ 
by $L^{-1/2}\, \exp(\mathrm{i}k_0 z)\,\delta_{nn'}\delta_{ss'}
 + g_{n's'}(z)$
 in Eqs.\ (\ref{expansion}) and (\ref{system}).
 Then, applying the Fourier transformation
 to \req{system} with $\bm{r}_0=\bm{r}_c$
 and treating $\tilde\psi_1$ as small perturbation,
 in the first approximation we obtain
\bea&&
   \left[ (\hbar^2/2\mu) \,(k^2-k_0^2) + E_{n's'}^\perp
   - E_{ns}^\perp \right] \tilde{g}_{n's'}(k)
%AA% \nonumber\\&&\qquad\qquad
   = - L^{-1}\,\tilde{V}_{ns,n's'}(r_c, k-k_0),
\label{system-Fourier}
\eea
where
\beq
   \tilde{V}_{ns,n's'}(r_c, k) =
   \int_{-\infty}^\infty \mathrm{e}^{-\mathrm{i} kz}
   \,V_{ns,n's'}(r_c,z)\,\dd z.
% \label{V-Fourier}
\eeq
Using Eqs.\ (A3)---(A10) of \citet{P94}, we can convert
$\tilde{V}_{ns,n's'}(r_c, k)$ into
\beq
   \tilde{V}_{ns,n's'}(r_c, k) = - {e^2}\,
   \tilde{v}_{ns,n's'}\left(\frac{r_c}{\am\,\sqrt{2}}, \sqrt{2}\,\am k\right),
\label{V-v}
\eeq
where
%AA% \bea
%AA% &&\hspace*{-.7em}
        \beq
   \tilde{v}_{ns,n's'}(\rho,\varkappa) = (-1)^{(|s|-s+|s'|-s')/2}
   \sum_{l=0}^{n_\rho + n'_\rho} (-1)^l
%AA% \nonumber\\&&\hspace{.5em}\times\hspace*{-1.5em}
   \sum_{m=\mathrm{max}(0,l-n'_\rho)}^{\mathrm{min}(n_\rho,l)}
   \hspace*{-1em}
   a_{m n_\rho |s|} a_{l-m, n'_\rho,|s'|} \sqrt{\tilde{s}!\,\tilde{s}'!}
   \,\tilde{v}_{0\tilde{s},0\tilde{s}'}(\rho,\varkappa),
\label{vnnss}
        \eeq
%AA% \eea
and, assuming $q\geq0$,
\bea
&&\hspace*{-.7em}
   \tilde{v}_{0,s,0,s+q}(\rho,\varkappa) = 
      \tilde{v}_{0,s+q,0,s}(\rho,\varkappa) = 
   \rho^q \sum_{m=0}^s a_{msq}\rho^{2m}
%E% \nonumber\\&&\hspace{-.7em}\times
   \int_0^1 % \hspace*{-.5em}
   \exp\left[-\frac{\varkappa^2}{4}\frac{1-t}{t} - \rho^2 t
   \right]
   t^{2m+q-1} (1-t)^{s-m} \,\dd t.
\label{vss}
\eea
In Eqs.\ (\ref{vnnss}) and (\ref{vss}), we have defined
\begin{subequations}
\bea
&&\hspace*{-.7em}
   n_\rho=n+ (s-|s|)/2,
\quad
   n'_\rho=n' + (s'-|s'|)/2,
\\&&\hspace*{-.7em}
   \tilde{s}=(|s|+s+|s'|-s')/2+l,
\\&&\hspace*{-.7em}
   \tilde{s}'=(|s|-s+|s'|+s')/2+l,
\\&&\hspace*{-.7em}
   a_{mns} = \frac{\sqrt{n!\,(n+s)!}}{m!\,(n-m)!\,(m+s)!} 
\eea
\end{subequations}
($n_\rho$ and $n'_\rho$ are the radial 
quantum numbers of the Landau functions ---
e.g., \citealp{LaLi-QM}).

For fixed quantum numbers $n_f,s_f$,
and a fixed sign of the $z$-projection of the relative momentum
$k_f$ of the final state, $\dd\nu_f$ in \req{dsigma}
equals $L\,\dd k_f/2\pi=(L/2\pi)\,(\mu/\hbar^2 |k_f|)\,\dd E_f$.
Therefore, the cross section of photoabsorption by an electron-proton 
pair with initial quantum numbers $n_i$ and $s_i$, longitudinal wave vector
$k_i$, and transverse pseudomomentum $K_\perp$ is
\beq
   \sigma(k_i, K_\perp, n_i, s_i, \omega) =
   \sum_{n_f,s_f,\mathrm{sign}k_f} 
   \frac{2\pi L \mu}{\hbar^2 |k_f| \omega\,c}\,
   \left|\bm{e}\cdot\langle f | \bm{j}_\mathrm{eff} | i \rangle
   \right|^2 .
\label{partial}
\eeq
Here, the sum is performed over those $n_f$ and $s_f$ which are
permitted by the energy conservation law,
\beq
   E_{n_f s_f}^\perp + \frac{\hbar^2 k_f^2}{2\mu}
   = E_{n_i s_i}^\perp + \frac{\hbar^2 k_i^2}{2\mu} + \hbar\omega .
\label{en-conserv}
\eeq
A general expression for 
$\bm{j}_\mathrm{eff}$ has been derived by 
\citet{PP97}. In the dipole approximation,
it reduces to
\beq
 \bm{j}_\mathrm{eff} = e \left(\frac{\bm{\pi}}{\mel}
 + \frac{\bm{\Pi}}{\mpr} \right),
\label{jeff}
\eeq
where $\bm{\pi}$ is defined by \req{pi}, and
\beq
   \bm{\Pi} = \bm{p}-\frac{e}{2c}\bm{B}\times\bm{r}.
\eeq
The circular components
of operators $\bm{\pi}$ and $\bm{\Pi}$,
$\pi_{\pm1}=(\pi_x\pm\mathrm{i}\pi_y)/\sqrt{2}$ 
and $\Pi_{\pm1}=(\Pi_x\pm\mathrm{i}\Pi_y)/\sqrt{2}$,
transform one Landau state $|n,s\rangle_\perp$,
characterized by the function $\Phi_{ns}(\bm{r}_\perp)$,
into another Landau state,\footnote{%
The square-root factors in Eqs.\ (\ref{Pi+}) and (\ref{Pi-})
were interchanged by mistake 
in corresponding Eq.\ (A3b) of \citet{PP97}.
}
\begin{subequations}
\label{pi+-}
\bea
   \pi_{+1} |n,s\rangle_\perp &=& - \frac{\mathrm{i}\hbar}{\am}
   \sqrt{n+1}\,|n+1,s-1\rangle_\perp,
\\
   \pi_{-1} |n,s\rangle_\perp &=&  \frac{\mathrm{i}\hbar}{\am}
   \sqrt{n}\,|n-1,s+1\rangle_\perp,
\\
   \Pi_{+1} |n,s\rangle_\perp &=& - \frac{\mathrm{i}\hbar}{\am}
   \sqrt{n+s}\,|n,s-1\rangle_\perp,
\label{Pi+}
\\
   \Pi_{-1} |n,s\rangle_\perp &=&  \frac{\mathrm{i}\hbar}{\am}
   \sqrt{n+s+1}\,|n,s+1\rangle_\perp.
\label{Pi-}
\eea
\end{subequations}
Since $\bm{e}\cdot\bm{j}=
e_{+1}\, j_{-1} + e_0\, j_0 + e_{-1}\, j_{+1}$,
the matrix element with $\pi_{+1}$ and $\Pi_{+1}$ 
contributes to $\sigma_{-1}$, and vice versa.

In the first Born approximation, 
using Eqs.\ (\ref{system-Fourier}), (\ref{en-conserv}), 
(\ref{jeff}),and (\ref{pi+-}), we obtain:
\begin{subequations}
\label{mxel}
\bea
&&\hspace*{-.7em}
   \langle f | \bm{j}_\mathrm{eff} | i \rangle_0
   = \frac{ e }{L\,\mu\,\omega}\,(k_i-k_f)\,
   \tilde{V}_{n_i s_i n_f s_f} ,
\\&&\hspace*{-.7em}
   \langle f | \bm{j}_\mathrm{eff} | i \rangle_{-1}
   = 
%AA% \nonumber\\&&
   - \frac{\mathrm{i} e}{ L \am} \bigg[
   \frac{ \sqrt{n_f + 1}\,\tilde{V}_{n_i s_i, n_f + 1, s_f-1}
   - \sqrt{n_i}\,\tilde{V}_{n_i-1, s_i+1, n_f s_f}
   }{ \mel\,(\omega + \omc)}
\nonumber\\&& +
   \frac{ \sqrt{n_f + s_f}\,\tilde{V}_{n_i s_i, n_f, s_f-1}
   - \sqrt{n_i+s_i+1}\,\tilde{V}_{n_i, s_i+1, n_f s_f}
   }{ \mpr\,(\omega - \omp)} \bigg],
\nonumber\\&&
\\&&\hspace*{-.7em}
   \langle f | \bm{j}_\mathrm{eff} | i \rangle_{+1}
   = 
%AA% \nonumber\\&& 
   \frac{\mathrm{i} e}{ L \am} \bigg[
   \frac{ \sqrt{n_f}\,\tilde{V}_{n_i s_i, n_f - 1, s_f+1}
   - \sqrt{n_i+1}\,\tilde{V}_{n_i+1, s_i-1, n_f s_f}
   }{ \mel\,(\omega - \omc)}
\nonumber\\&&\hspace*{-2pt} +
   \frac{ \sqrt{n_f + s_f+1}\,\tilde{V}_{n_i s_i, n_f, s_f+1}
   - \sqrt{n_i+s_i}\,\tilde{V}_{n_i, s_i-1, n_f s_f}
   }{ \mpr\,(\omega + \omp)} \bigg].
\nonumber\\&&
\eea
\end{subequations}
Here, for brevity, $\tilde{V}_{nsn's'}\equiv
\tilde{V}_{nsn's'}(r_c, k_f-k_i)$.

Equations (\ref{partial}) and (\ref{mxel}) 
provide the partial cross sections for one electron-proton pair
in a given state.
Provided there are $n_\mathrm{e}$ electrons per unit volume,
the number of electrons interacting with a given proton
and having $k_i$ in the interval $\dd k_i$
and $r_c$ in the surface element $\dd^2 r_c$
is 
\beq
  \dd N_i = n_\mathrm{e} L \,\dd^2 r_c\, \mathcal{F}_\|(k_i)\, \dd k_i
  \,\mathcal{F}_{n_i s_i}.
\eeq
Here,
\beq
   \mathcal{F}_\|(k_i) = \hbar\, (2\pi\mu\kB T)^{-1/2}\,
   \exp(-\hbar^2 k_i^2/2\mu\kB T)
\eeq
is the Maxwell distribution of the continuum states over $k_i$, and
\beq
   \mathcal{F}_{n_i s_i} = (1-\mathrm{e}^{-\beta_\mathrm{e}})\,
   (1-\mathrm{e}^{-\beta_\mathrm{p}})\,
   \exp\left[ -n_i \beta_\mathrm{e} - (n_i+s_i) \beta_\mathrm{p}
   \right] 
\eeq
is the Boltzmann distribution over $n_i\geq0$ and $s_i \geq -n_i$.
Thus the total cross section is
\bea
   \sigma(\omega,T,B) & = & n_\mathrm{e} L \pi \am^4
   \sum_{n_is_i} \mathcal{F}_{n_i s_i}
   \int_{-\infty}^\infty \mathcal{F}_\|(k_i) \,\dd k_i
%AA% \nonumber\\&&\times
   \int_0^\infty K_\perp \dd K_\perp 
   \sigma(k_i, K_\perp, n_i, s_i, \omega).
\label{sigma-int}
\eea
For every polarization, let us write $\sigma(\omega,T,B)$ 
in the form of \req{sigma-e}. Then 
\bea
   \Lambda_\alpha^\mathrm{e} &=& \frac34 \sum_{n_is_i} \mathcal{F}_{n_i s_i}
   \sum_{n_f,s_f,\mathrm{sign}\varkappa_f} 
   \int_0^\infty\frac{\dd\varkappa_i}{|\varkappa_f|}
%AA% \nonumber\\&&\times
   \exp[-(\beta_\mathrm{e}+\beta_\mathrm{p})\,\varkappa_i^2 /4]
   \,w_{n_i s_i n_f s_f}^\alpha(\varkappa_f - \varkappa_i),
\label{Lambda-e}
\eea
where, taking into account \req{en-conserv}, we have
\beq
   \varkappa_f^2 = \varkappa_i^2 + 4(n_i-n_f) + 4\frac{\mel}{\mH}
   (s_i-s_f) +4\frac{\mpr}{\mH}\,\frac{u}{\beta_\mathrm{e}},
\eeq
and
the functions $w_{n_i s_i n_f s_f}^\alpha(\varkappa_f - \varkappa_i)$
in the integrand are defined
according to Eqs.\ (\ref{V-v}), (\ref{mxel}), 
and (\ref{sigma-int}) as
\begin{subequations}
\label{w}
\bea
&&\hspace*{-.7em}
   w_{nsn's'}^{0}(\varkappa) = \varkappa^2
   \int_0^\infty \rho\,\dd\rho\,
   \big| \tilde{v}_{nsn's'}(\rho,\varkappa) \big|^2,
\\&&\hspace*{-.7em}
   w_{nsn's'}^{-1}(\varkappa) = \frac{2}{(1+\mel/\mpr)^2}
   \int_0^\infty \rho\,\dd\rho\,
%AA% \nonumber\\&&\times
   \Big| \sqrt{n'}\,\tilde{v}_{ns,n'-1,s'+1}(\rho,\varkappa)
   - \sqrt{n+1}\,\tilde{v}_{n+1,s-1,n's'}(\rho,\varkappa)
\nonumber\\&&\qquad
   +\frac{\mel}{\mpr}\,\frac{\omega-\omc}{\omega+\omp}\,
   \big[ \sqrt{n'+s'+1}\,\tilde{v}_{nsn',s'+1}(\rho,\varkappa)
%AA% \nonumber\\&&\qquad\qquad
   - \sqrt{n+s}\,\tilde{v}_{n,s-1,n's'}(\rho,\varkappa) 
   \big] \Big|^2,
\label{w-}
\\&&\hspace*{-.7em}
   w_{nsn's'}^{+1}(\varkappa) = \frac{2}{(1+\mel/\mpr)^2}
   \int_0^\infty \rho\,\dd\rho\,
%AA% \nonumber\\&&\times
   \Big| \sqrt{n'+1}\,\tilde{v}_{ns,n'+1,s'-1}(\rho,\varkappa)
   - \sqrt{n}\,\tilde{v}_{n-1,s+1,n's'}(\rho,\varkappa)
\nonumber\\&&\qquad
   +\frac{\mel}{\mpr}\,\frac{\omega+\omc}{\omega-\omp}\,
   \big[ \sqrt{n'+s'}\,\tilde{v}_{nsn',s'-1}(\rho,\varkappa)
%AA% \nonumber\\&&\qquad\qquad
   - \sqrt{n+s+1}\,\tilde{v}_{n,s+1,n's'}(\rho,\varkappa) 
   \big] \Big|^2.
\label{w+}
\eea
\end{subequations}

\end{appendix}

\end{document}